\documentclass[aps,pra,twocolumn,showpacs,floatfix]{revtex4}
\usepackage{graphicx}
\begin{document}
\title{Ultracold atom-molecule collisions and bound states in
magnetic fields: \\ tuning zero-energy Feshbach resonances in
He-NH ($^3\Sigma^-$)}
\author{Maykel Leonardo Gonz{\'a}lez-Mart{\'\i}nez}
\affiliation{Departamento de F{\'\i}sica General y
Matem{\'a}ticas, InSTEC, Habana 6163, Cuba}
\author{Jeremy M. Hutson}
\affiliation{Department of Chemistry, University of Durham, South
Road, Durham, DH1~3LE, England}

\date{\today: arXiv:physics/0610214}

\begin{abstract}
We have generalized the BOUND and MOLSCAT packages to allow
calculations in basis sets where the monomer Hamiltonians are
off-diagonal and used the new capability to carry out bound-state
and scattering calculations on $^3$He-NH and $^4$He-NH as a
function of magnetic field. Following the bound-state energies to
the point where they cross thresholds gives very precise
predictions of the magnetic fields at which zero-energy Feshbach
resonances occur. We have used this to locate and characterize two
very narrow Feshbach resonances in $^3$He-NH. Such resonances can
be used to tune elastic and inelastic collision cross sections,
and sweeping the magnetic field across them will allow a form of
quantum control in which separated atoms and molecules are
associated to form complexes. For the first resonance, where only
elastic scattering is possible, the scattering length shows a pole
as a function of magnetic field and there is a very large peak in
the elastic cross section. For the second resonance, however,
inelastic scattering is also possible. In this case the pole in
the scattering length is dramatically suppressed and the cross
sections show relatively small peaks. The peak suppression is
expected to be even larger in systems with stronger inelasticity.
The results suggest that calculations on ultracold molecular
inelastic collisions may be much less sensitive to details of the
potential energy surface than has been believed.
\end{abstract}


\maketitle

\section{Introduction}

Over the last 5 years, it has become possible to control the
behavior of ultracold atomic gases by tuning the interactions
between atoms using applied magnetic fields
\cite{Hutson:IRPC:2006, Koehler:RMP:2006}. Notable successes have
included the controlled implosion of Bose-Einstein condensates
\cite{Roberts:2001} and the production of molecules in both
bosonic \cite{Donley:2002, Herbig:2003, Xu:2003,
Durr:mol87Rb:2004} and fermionic \cite{Regal:40K2:2003,
Strecker:2003, Cubizolles:2003, Jochim:Li2pure:2003} quantum
gases. Long-lived molecular Bose-Einstein condensates of fermion
dimers have been produced \cite{Jochim:Li2BEC:2003,
Zwierlein:2003, Greiner:2003}, and the first signatures of
ultracold triatomic \cite{Kraemer:2006} and tetraatomic
\cite{Chin:2005} molecules have been observed. The new
capabilities in atomic physics have had important applications in
other areas: for example, the tunability of atomic interactions
has allowed exploration of the crossover between Bose-Einstein
condensation (BEC) and Bardeen-Cooper-Schrieffer (BCS) behavior in
dilute gases \cite{Bartenstein:crossover:2004,
Regal:res-cond:2004, Zwierlein:2004}.

In parallel with the work on atomic gases, there have been intense
efforts to cool molecules directly from high temperature to the
ultracold regime. Molecules such as NH$_3$, OH and NH have been
cooled from room temperature to the milliKelvin regime by a
variety of methods including buffer-gas cooling
\cite{Weinstein:CaH:1998, Egorov:2004} and Stark deceleration
\cite{Bethlem:IRPC:2003, Bethlem:2006}. Directly cooled molecules
have been successfully trapped at temperatures around 10 mK, and
there are a variety of proposals for ways to cool them further,
including evaporative cooling, sympathetic cooling and
cavity-assisted cooling \cite{Domokos:2002, Chan:2003}.

The possibility of controlling {\it molecular} interactions in the
same way as {\it atomic} interactions is of great interest.
Several groups have begun to explore the effects of external
fields on ultracold molecular collisions \cite{Krems:IRPC:2005}.
Volpi and Bohn \cite{Volpi:2002} investigated collisions of
$^{17}$O$_2$ with He and found a very strong enhancement of
spin-flipping cross sections even for weak magnetic fields. Krems
{\em et al.}\ \cite{Krems:henh:2003} and Cybulski {\em et al.}\
\cite{Cybulski:2005} investigated spin-flipping collisions of NH
with He and found a similar dependence on magnetic field. Krems
and Dalgarno \cite{Krems:mfield:2004, Krems:FWQC:2004} elaborated
the formal theory of scattering in a magnetic field for a variety
of atom-molecule and molecule-molecule cases involving molecules
in $^2\Sigma$ and $^3\Sigma$ states. Ticknor and Bohn
\cite{Ticknor:OHmag:2005} investigated OH+OH collisions and found
that in this case magnetic fields could {\it suppress} inelastic
collisions. Lara {\em et al.}\ \cite{Lara:PRL:2006, Lara:PRA:2006}
investigated the very complicated case of OH + Rb collisions,
using basis sets designed to allow a magnetic field to be applied,
though their initial calculations were for zero field.

The effects of electric fields have also been investigated.
Avdeenkov and Bohn \cite{Avdeenkov:2002, Avdeenkov:2003}
investigated OH-OH collisions in the presence of electric fields
that caused alignment of the molecules. They identified novel
field-linked states arising from long-range avoided crossings
between effective potential curves in the presence of a field
\cite{Avdeenkov:2003, Avdeenkov:2004, Avdeenkov:2005,
Ticknor:long-range:2005}. Avdeenkov {\em et al.}\
\cite{Avdeenkov:2006} have also investigated the effects of very
high electric fields on collisions of closed-shell molecules. Very
recently, Tscherbul and Krems \cite{Tscherbul:2006} have explored
the effect of combined electric and magnetic fields on He-CaH
collisions and observed significant suppression of spin-flipping
transitions at high electric fields in the cold regime
($\sim0.5$~K).

An important technique used to produce dimers in ultracold atomic
gases is magnetic tuning across Feshbach resonances
\cite{Hutson:IRPC:2006, Koehler:RMP:2006}. A Feshbach resonance
\cite{Feshbach:1958, Feshbach:1962} occurs whenever a bound state
associated with one potential curve lies above the threshold for
another curve. The resonance thus corresponds to a level embedded
in a continuum, which is a quasibound state. In the atomic case,
the thresholds that produce low-energy Feshbach resonances are
associated with different hyperfine states of the interacting
atoms. It is often possible to tune a resonance across threshold
(from above or below) by applying a magnetic field. This produces
an avoided crossing between atomic and molecular states. If the
magnetic field is tuned across the resonance slowly enough to
follow the avoided crossing adiabatically, pairs of atoms can be
converted into molecules or {\em vice versa}.

Molecules have a much richer energy level structure than atoms,
and there are many additional types of Feshbach resonance. In
particular, {\it rotational} Feshbach resonances can occur
\cite{Ashton:1983, Hutson:HDAr:1983} and have significant
influences on ultracold molecular collisions
\cite{Bohn:fesh:2002}. Other small energy level splittings, such
as spin-rotation and $\Lambda$-doubling, can also cause resonances
in molecular scattering. It is of great interest to characterize
such resonances and their field-dependence, both to understand
their influence on collision cross sections and to prepare the
ground for experiments that associate molecules by Feshbach
resonance tuning.

The NH molecule is particularly topical in cold and ultracold
molecule studies. It is a dipolar molecule with a $^3\Sigma^-$
ground state, so it is both electrostatically and magnetically
trappable. It has been cooled by beam-loaded buffer-gas cooling
\cite{Egorov:2004} and is a promising candidate for molecular beam
deceleration and trapping \cite{vandeMeerakker:2003}. Krems {\em
et al.}\ \cite{Krems:henh:2003} and Cybulski {\em et al.}\
\cite{Cybulski:2005} have calculated potential energy surfaces for
He-NH and used them in scattering calculations. Cybulski {\em et
al.}\ also calculated zero-field bound states of the He-NH Van der
Waals complex. An electronic excitation spectrum of the Van der
Waals complex has been observed by Kerenskaya {\em et al.}\
\cite{Kerenskaya:2004}. Sold\'an and Hutson \cite{Soldan:2004}
have investigated the interaction potentials for NH with Rb, and
Dhont {\em et al.}\ \cite{Dhont:2005} have developed interaction
potential energy surfaces for NH-NH in the singlet, triplet and
quintet states.

In the present article we describe the first calculations of the
bound states of a Van der Waals complex in a magnetic field. We
show how such calculations can be used to locate zero-energy
Feshbach resonances as a function of applied field. Elastic and
inelastic cross sections can then be tuned by sweeping the field
across the Feshbach resonance. Such field sweeps could also be
used to transfer unbound atom-molecule or molecule-molecule pairs
into bound states of the corresponding complex.

A remarkable conclusion of the present paper is that, in the
presence of inelastic scattering, the poles in scattering lengths
that characterize low-energy Feshbach resonances in atomic systems
\cite{Koehler:RMP:2006} are dramatically suppressed and elastic
and inelastic cross sections show relatively small peaks as
resonances cross thresholds. The numerical results obtained here
allow us to test analytical formulae for this effect recently
given by Hutson \cite{Hutson:res:2006}.

\section{Methods for bound-state calculations}

We consider the case of an NH molecule interacting with a He atom
in the presence of a magnetic field. The Hamiltonian for this in
Jacobi coordinates $(R,\theta)$ is
\begin{equation}
{\hat H}=-\frac{\hbar^2}{2\mu} R^{-1} \frac{d^2}{dR^2}R +
\frac{\hat L^2}{2\mu R^2} + {\hat H}_{\rm mon} + {\hat H}_{\rm Z}
+ V(R,\theta) , \label{eqham}
\end{equation} where $\hat L^2$ is
the space-fixed operator for end-over-end rotation, ${\hat H}_{\rm
mon}$ is the Hamiltonian for the NH monomer, ${\hat H}_{\rm Z}$ is
the Zeeman interaction and $V(R,\theta)$ is the intermolecular
potential. For simplicity we consider the NH molecule to be a
rigid rotor, but the generalization to include NH vibrations is
straightforward. The NH monomer Hamiltonian is therefore
\begin{equation}
{\hat H}_{\rm mon} = \hbar^{-2} b_{\rm NH} \hat N^2 + {\hat
H}_{\rm SN} + {\hat H}_{\rm SS},
\end{equation}
where $b_{\rm NH}=16.343\;{\rm cm}^{-1}$ is the rotational
constant of NH in its ground vibrational level
\cite{Brazier:1986},
\begin{equation}
{\hat H}_{\rm SN}=\gamma\hat N \cdot \hat S
\end{equation}
is the spin-rotation operator, and
\begin{equation}
{\hat H}_{\rm SS}=\frac{2}{3}\lambda_{\rm SS}
\left[\frac{4\pi}{5}\right]^{\frac{1}{2}}\sqrt{6} \sum_q (-1)^q
Y_{2-q}(\hat r)\left[S\otimes S\right]^{(2)}_q
\end{equation}
is the spin-spin operator written in space-fixed coordinates
\cite{Mizushima}. $\hat N$ and $\hat S$ are the operators for the
rotational and spin angular momenta. The numerical values for the
spin-rotation and spin-spin constants are $\gamma = -0.0055\;{\rm
cm}^{-1}$ and $\lambda_{\rm SS}=0.920\;{\rm cm}^{-1}$
\cite{Mizushima}.

There are several basis sets that could be used to expand the
eigenfunctions of Eq.\ (\ref{eqham}). We consider two of them in
the present work, which we refer to as the {\it coupled} and {\it
uncoupled} basis sets. Both basis sets represent the end-over-end
rotation with quantum numbers $|LM_L\rangle$, where $L$ is a
rotational quantum number and $M_L$ is its projection onto the
space-fixed $Z$ axis.

We use the convention that quantum numbers that describe a {\it
monomer} are represented with lower-case letters, and reserve
capital letters to describe states of the complex as a whole. In
the absence of a magnetic field (or a perturbing atom), the
rotational states of NH are approximately described by quantum
numbers $n$, $s$ and $j$, where $n$ represents the mechanical
rotational of NH, $s$ is the electron spin, and $j$ is the vector
sum of $n$ and $s$. In the coupled representation for the He-NH
problem, we use basis functions $|nsjm_j\rangle |LM_L\rangle$ that
retain these monomer quantum numbers, with $m_j$ the projection of
$j$ onto the space-fixed $Z$ axis. In the uncoupled
representation, we use instead basis functions $|nm_n\rangle
|sm_s\rangle |LM_L\rangle$, where $m_n$ and $m_s$ are the
projections of $n$ and $s$ individually.

In both basis sets we use, the matrix elements of $\hat L^2$ are
diagonal in all quantum numbers and are simply $\hbar^2L(L+1)$.
The rotational part of the monomer Hamiltonian is also diagonal,
with matrix elements $b_{\rm NH} n(n+1)$. The remaining matrix
elements of the NH monomer Hamiltonian in the two basis sets are

\begin{widetext}
\begin{eqnarray}
\langle nsjm_j| {\hat H}_{\rm SN} |n'sj'm_j'\rangle &=& \delta_{n
n'}\delta_{j j'}\delta_{m_j m_j'} \gamma (-1)^{n+j+s}
\left[n(n+1)(2n+1)s(s+1)(2s+1)\right]^{\frac{1}{2}}
\left\{\begin{array}{ccc} s & n & j \\ n & s & 1
\end{array}\right\};
\\
\langle nsjm_j| {\hat H}_{\rm SS} |n'sj'm_j'\rangle &=& \delta_{j
j'}\delta_{m_j m_j'} \frac{2\sqrt{30}}{3}\lambda_{\rm SS}
(-1)^{j+n'+n+s} \left[(2n+1)(2n'+1)\right]^{\frac{1}{2}}
\left(\begin{array}{ccc} n & 2 & n' \\ 0 & 0 & 0
\end{array}\right) \left\{\begin{array}{ccc} s & n' & j \\ n & s &
2 \end{array}\right\}
\end{eqnarray}
and
\begin{eqnarray}
\langle sm_s| \langle nm_n| {\hat H}_{\rm SN} |n'm_n'\rangle
|sm_s'\rangle &=& \delta_{n n'} \delta_{m_n m_n'} \delta_{m_s
m_s'} \gamma m_n m_s \nonumber \\ &+& \left(\delta_{n
n'}\delta_{m_n m_n' \pm 1}\delta_{m_s m_s'\mp 1}\right)
\frac{\gamma}{2} \left[n(n+1)-{m_n'}(m_n' \pm
1)\right]^{\frac{1}{2}} \left[s(s+1)-{m_s'}(m_s' \mp
1)\right]^{\frac{1}{2}};
\\
\langle sm_s| \langle nm_n| {\hat H}_{\rm SS} |n'm_n'\rangle
|sm_s'\rangle &=& \frac{2\sqrt{30}}{3}\lambda_{\rm
SS}(-1)^{s-m_s-m_n}
\left[(2n+1)(2n'+1)\right]^{\frac{1}{2}}\left[s(s+1)(2s+1)\right]
\left(\begin{array}{ccc} n & 2 & n' \\ 0 & 0 & 0
\end{array}\right) \nonumber \\* &&\times
\left\{\begin{array}{ccc} 1 & 1 & 2 \\ s & s & s
\end{array}\right\} \sum_q (-1)^q \left(\begin{array}{ccc} n & 2 &
n' \\ -m_n & -q & m_n' \end{array}\right) \left(\begin{array}{ccc}
s & 2 & s \\ -m_s & q & m_s' \end{array}\right).
\end{eqnarray}
It may be noted that ${\hat H_{\rm mon}}$ is approximately
diagonal in the coupled representation but not in the uncoupled
representation. In the coupled representation, the only
off-diagonal terms are matrix elements of ${\hat H_{\rm SS}}$ that
couple different rotational states with $\Delta n=\pm2$.

The Zeeman Hamiltonian for NH, neglecting rotational and
anisotropic spin terms \cite{Brown:p646} is
\begin{equation}
{\hat H}_{\rm Z} = g_{\rm e}\mu_{\rm B}{\hat B} \cdot {\hat S},
\end{equation}
where $g_{\rm e}$ is the $g$-factor for the electron, $\mu_{\rm
B}$ the Bohr magneton and $\hat B$ is the magnetic field vector.
The matrix elements of this operator are
\begin{eqnarray}
\langle nsjm_j| {\hat H}_{\rm Z} |n'sj'm_j'\rangle &=& \delta_{n
n'}\delta_{m_j m_j'} g_{\rm e} \mu_{\rm B} B (-1)^{n+s-m_j+1}
\left[s(s+1)(2s+1)(2j+1)(2j'+1)\right]^{\frac{1}{2}} \nonumber \\*
&& \times \left(\begin{array}{ccc} j & 1 & j' \\ -m_j & 0 & m_j
\end{array}\right) \left\{\begin{array}{ccc} s & j' & n \\ j & s &
1 \end{array}\right\}
\end{eqnarray}
and
\begin{equation}
\langle sm_s| \langle nm_n| {\hat H}_{\rm Z} |n'm_n'\rangle
|sm_s'\rangle = \delta_{n n'}\delta_{m_n m_n'}\delta_{m_s m_s'}
g_{\rm e}\mu_{\rm B} B m_s,
\end{equation}
where the magnetic field direction has been chosen as the $Z$ axis
and $B$ is the field strength. This is diagonal in the uncoupled
representation but not in the coupled representation.

The intermolecular potential is conveniently expanded in Legendre
polynomials,
\begin{equation} V(R,\theta) = \sum_\lambda V_\lambda(R)
P_\lambda(\cos\theta). \end{equation} The matrix elements of the
Legendre polynomials in the coupled and uncoupled basis sets are
\begin{eqnarray}
\langle LM_L| \langle nsjm_j| P_\lambda(\cos\theta)
|n'sj'm_j'\rangle |L'M_L'\rangle &=&
\left[(2n+1)(2n'+1)(2j+1)(2j'+1)(2L+1)(2L'+1)\right]^{\frac{1}{2}}
\nonumber \\* && \times \left(\begin{array}{ccc} n & \lambda & n'
\\ 0 & 0 & 0 \end{array}\right) \left(\begin{array}{ccc} L &
\lambda & L' \\ 0 & 0 & 0 \end{array}\right)
\sum_{m_\lambda}(-1)^{s+j+j'+\lambda+m_{\lambda}-M_L-m_j}
\nonumber \\* && \times \left(\begin{array}{ccc} L & \lambda & L'
\\ -M_L & -m_{\lambda} & M_L' \end{array}\right)
\left(\begin{array}{ccc} j & \lambda & j' \\ -m_j & m_{\lambda} &
m_j' \end{array}\right) \left\{\begin{array}{ccc} j & j' & \lambda
\\ n' & n & s \end{array}\right\}
\end{eqnarray}
and
\begin{eqnarray}
\langle LM_L| \langle sm_s| \langle nm_n| P_\lambda(\cos\theta)
|n'm_n'\rangle |sm_s'\rangle |L'M_L'\rangle &=& \delta_{m_s m_s'}
\left[(2n+1)(2n'+1)(2L+1)(2L'+1)\right]^{\frac{1}{2}} \nonumber
\\* && \times \left(\begin{array}{ccc} n & \lambda & n' \\ 0 & 0 &
0 \end{array}\right) \left(\begin{array}{ccc} L & \lambda & L' \\
0 & 0 & 0 \end{array}\right)
\sum_{m_\lambda}(-1)^{m_{\lambda}-M_L-m_n} \nonumber \\* && \times
\left(\begin{array}{ccc} L & \lambda & L' \\ -M_L & -m_{\lambda} &
M_L' \end{array}\right) \left(\begin{array}{ccc} n & \lambda & n'
\\ -m_n & m_{\lambda} & m_n' \end{array}\right).
\end{eqnarray}
This is off-diagonal in both representations.

We solve the bound-state Hamiltonian by a coupled channel method
\cite{Hutson:CPC:1994}. Denoting a complete set of channel quantum
numbers $(n,s,j,m_j,L,M_L)$ or $(n,m_n,s,m_s,L,M_L)$ by $i$, we
expand the total wavefunction
\begin{equation}
\Psi = R^{-1} \sum_i \Phi_i(\hat R,\hat r) \chi_i(R),
\label{eqfull} \end{equation} where $(\hat R,\hat r)$ represents
all coordinates {\it except} the intermolecular distance $R$ and
the {\it channel functions} $\Phi_i(\hat R,\hat r)$ are the
corresponding basis functions of the coupled or uncoupled basis
sets. Substituting this expansion into the total Schr\"odinger
equation yields a set of coupled differential equations for the
{\it radial functions} $\chi_i(R)$,
\begin{equation}
\left[-\frac{\hbar^2}{2\mu} \frac{d^2}{dR^2} + \frac{\hbar^2
L(L+1)}{2\mu R^2} - E \right]\chi_{i}(R)=-\sum_{i'} \biggl[\langle
i | {\hat H}_{\rm mon} + {\hat H}_{\rm Z}  + V(R,\theta) | i'
\rangle \biggr] \chi_{i'}(R).
\end{equation}
\end{widetext}
In the present work we solve the coupled equations to find bound
states using the BOUND program \cite{Hutson:bound:1993}, which
uses the algorithms described in ref.\
\onlinecite{Hutson:CPC:1994}. For $N$ channels there are $N$
coupled equations. However, it is actually necessary to propagate
a set of $N$ linearly independent solutions, so $\chi(R)$ is an
$N\times N$ matrix. The log-derivative matrix
$Y=(d\chi/dR)\chi^{-1}$ is propagated outwards from $R_{\rm min}$
and inwards from $R_{\rm max}$ to a common matching point $R_{\rm
mid}$ in the classically allowed region of the potential using
Johnson's algorithm \cite{Johnson:1973}. This is done for a series
of trial energies. If the energy is an eigenvalue of the
Hamiltonian, the determinant of the log-derivative matching matrix
$Y_{\rm match}=Y_{\rm in}-Y_{\rm out}$ is zero. Bound states are
located by searching for zeroes of {\it eigenvalues} of $Y_{\rm
match}$ as a function of energy. The log-derivative method
provides a generalised node count \cite{Johnson:1978,
Hutson:CPC:1994} which increases by one at each energy eigenvalue,
and this allows us to use bisection to identify regions of energy
that contain a bound state. The actual convergence on an energy
eigenvalue uses the secant method, which gives quadratic
convergence.

Version 5 of the BOUND program \cite{Hutson:bound:1993} contained
an interface to allow new basis sets to be added, but could only
handle basis sets in which the monomer Hamiltonian was diagonal.
We have extended the program to remove this restriction and
implemented the coupled and uncoupled basis sets described above.
We have also built in new loops over external fields to simplify
calculations on Stark and Zeeman effects.

\section{Results of bound-state calculations}

We have carried out bound-state calculations on $^4$He-NH and
$^3$He-NH using the potential energy surface of Krems {\em et
al.}\ \cite{Krems:henh:2003} (described in more detail as
potential 2 of Cybulski {\em et al.}\ \cite{Cybulski:2005}). The
basis set included all functions with $n \le 8$ and $L \le 7$. The
coupled equations were propagated outwards from 1.8 \AA\ to 3.57
\AA\ and inwards from 16.0 \AA\ to 3.57 \AA\ using a
log-derivative sector size of 0.025 \AA. WKB boundary conditions
were applied in each channel at $R_{\rm max}$ to improve the
convergence.

In zero field, the levels of He-NH are characterized by the total
angular momentum $\cal J$, which is the vector sum of $j$ and $L$.
The total parity is also conserved and is given by $(-1)^{n+L+1}$.
Since the lowest levels of NH have $n=0$, $j=s=1$ and ${\cal J} =
L$, $L\pm1$. The three different ${\cal J}$ levels corresponding
to each value of $L$ are very close together: the separation is
only about 10$^{-3}$~cm$^{-1}$ for $L=1$. The angular momentum
coupling scheme corresponds to case (B) of Dubernet, Flower and
Hutson \cite{Dubernet:1991}.

The field-free energies of He-NH on this potential have been
calculated previously by Cybulski {\em et al.}\
\cite{Cybulski:2005}. Our results agree with theirs to
$\pm10^{-4}$ cm$^{-1}$ for the levels with $L=0$ to 2, but are
approximately 0.0014 cm$^{-1}$ lower for the $L=3$ levels of
$^4$He-NH, which are bound by only 0.765 cm$^{-1}$. We attribute
the difference to lack of convergence of the radial basis set used
in their calculations. The results obtained from our program with
the coupled and uncoupled basis sets are identical to $\pm
10^{-9}$ cm$^{-1}$, which confirms the correctness of the code.

\begin{figure}[tb]
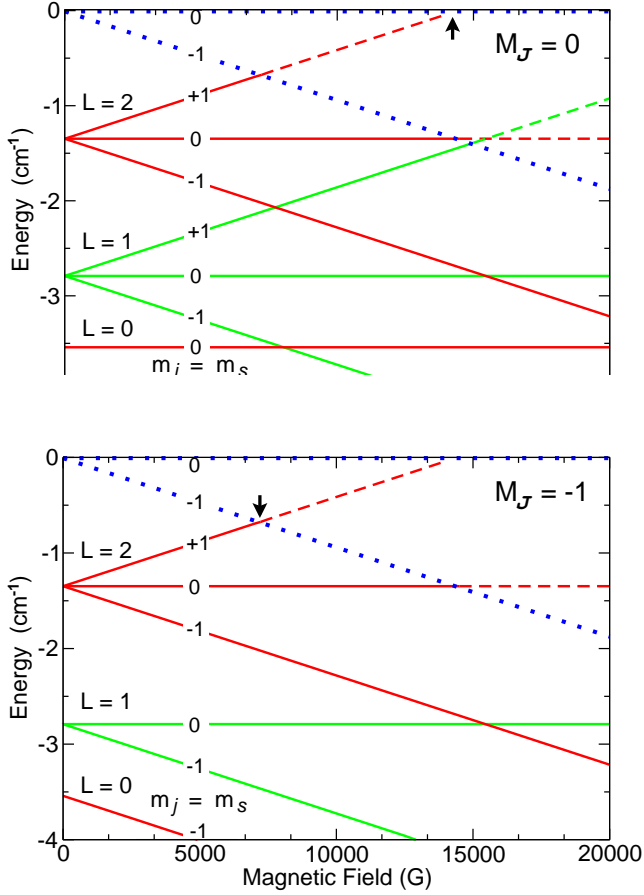

\includegraphics[width=85mm]{fig_BdepHe3M0.eps}
\includegraphics[width=85mm]{fig_BdepHe3M-1.eps}
\caption{Bound-state energy levels for $^3$He-NH for $M_{\cal
J}=0$ (upper panel) and $M_{\cal J}=-1$ (lower panel) as a
function of magnetic field $B$. Levels of odd parity are shown in
red and levels of even parity in green. Dissociation thresholds
are shown as dotted blue lines and quasibound levels are shown as
dashed lines. The arrows show the positions at which levels cross
$L=0$ thresholds.} \label{figbs3}
\end{figure}

\begin{figure}[tb]
\includegraphics[width=85mm]{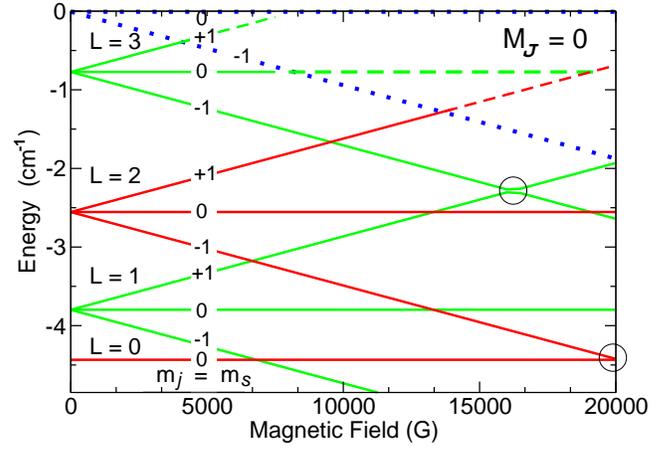}
\caption{Bound-state energy levels for $^4$He-NH for $M_{\cal
J}=0$ as a function of magnetic field $B$. Levels of odd parity
are shown in red and levels of even parity in green. Dissociation
thresholds are shown as dotted blue lines and quasibound levels
are shown as dashed lines. Avoided crossings are circled.}
\label{figbs4}
\end{figure}

In the presence of a field, $\cal J$ is very quickly destroyed.
The only rigorously good quantum numbers in a magnetic field are
parity and $M_{\cal J} = m_j + M_L = m_n + m_s + M_L$. The
bound-state energies for levels correlating with $n=0$ for
$^3$He-NH with $M_{\cal J}=0$ and $-1$ are shown in Fig.\
\ref{figbs3}. Each level splits into components that can be
labelled with the approximate quantum numbers $m_n=0$ and $m_s=0$,
$\pm1$. $L$ remains an essentially good quantum number except in
the vicinity of the avoided crossings. The major difference
between $M_{\cal J}$ values is that some $m_s$ levels are missing
for $M_{\cal J}\ne0$, but there are also small shifts of all the
energy levels. A similar plot for $^4$He-NH with $M_{\cal J}=0$ is
shown in Fig.\ \ref{figbs4}.

\begin{figure}[tb]
\includegraphics[width=85mm]{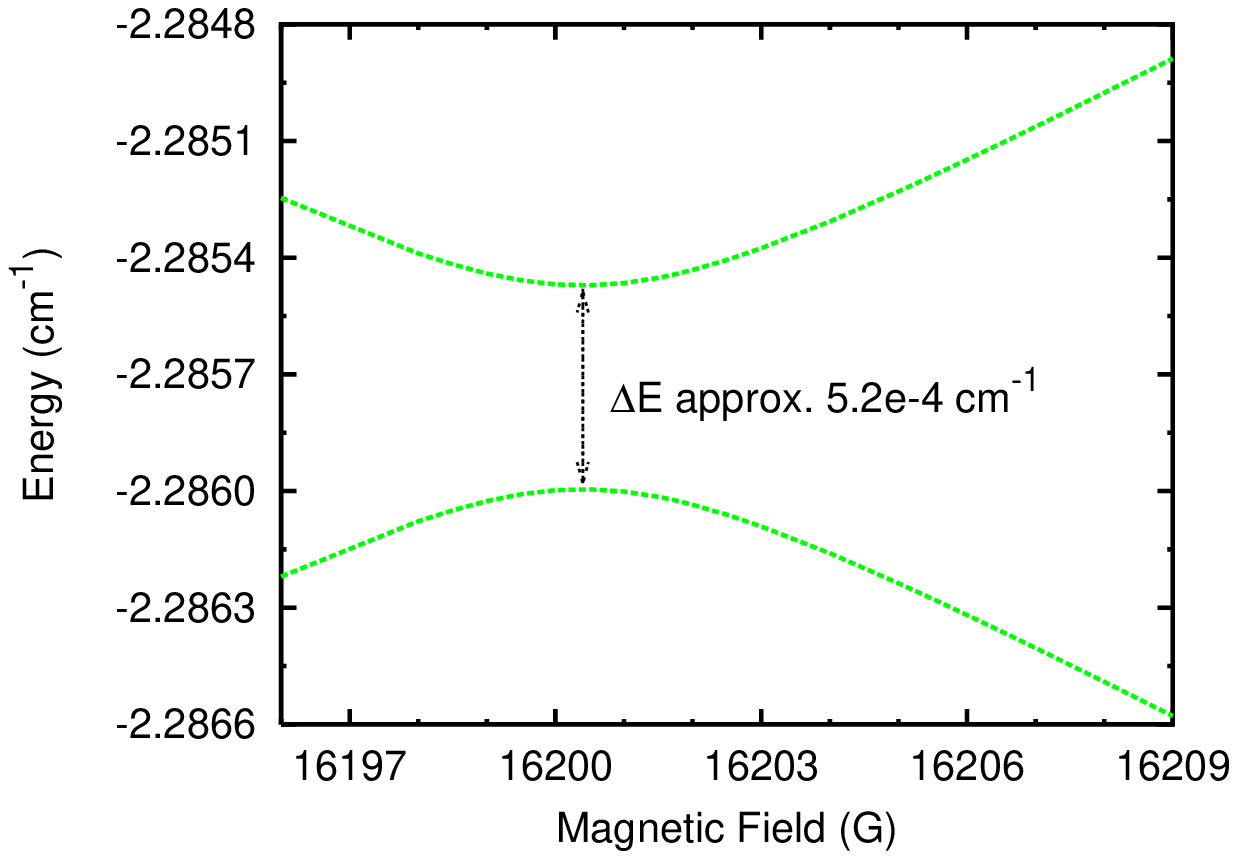}
\includegraphics[width=85mm]{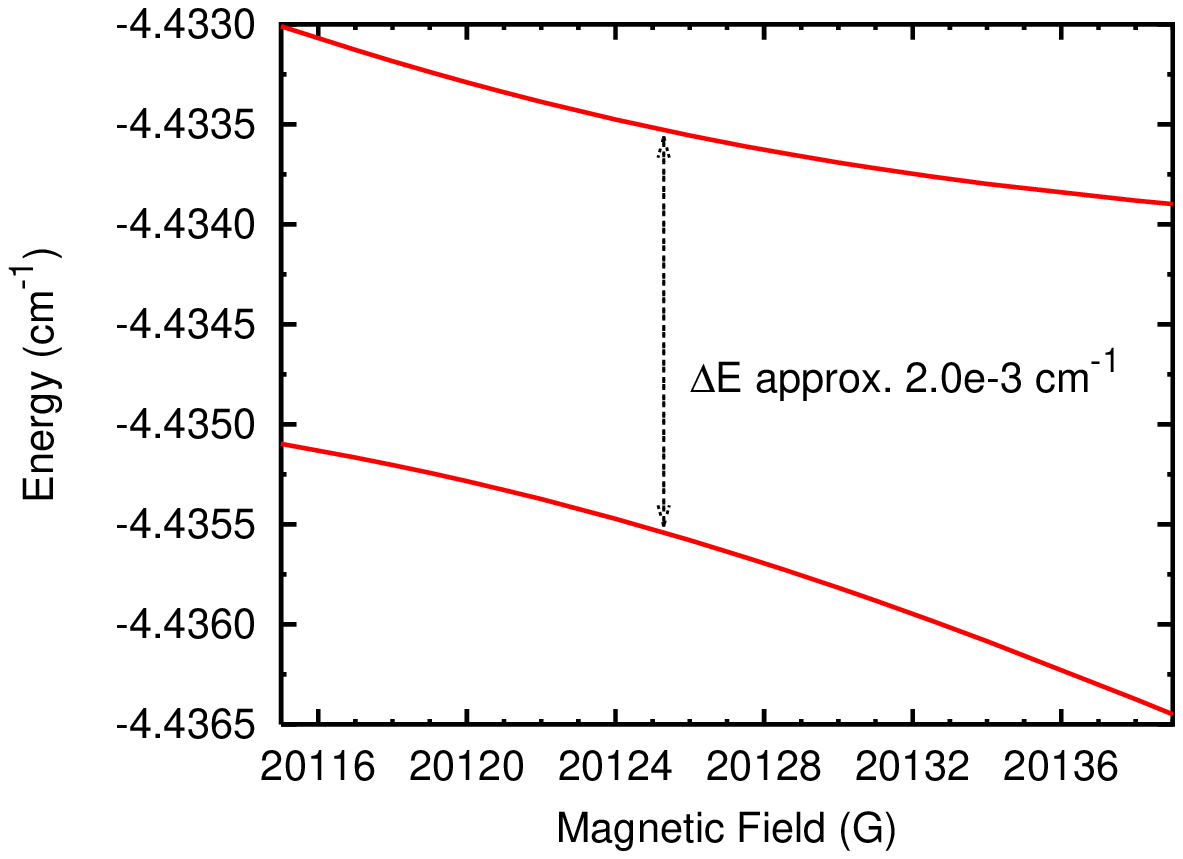}
\caption{Avoided crossing between levels with $m_s=+1$ and
$m_s=-1$ (upper panel) and between levels with $m_s=0$ and
$m_s=-1$ (lower panel) for $^4$He-NH as a function of magnetic
field $B$.} \label{figac4}
\end{figure}

\begin{figure}[tb]
\includegraphics[width=85mm]{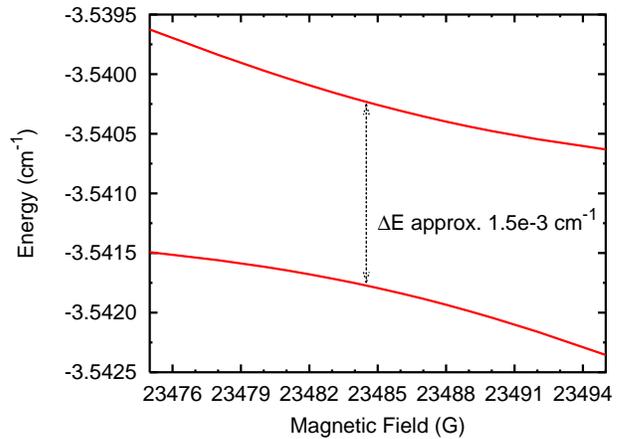}
\caption{Avoided crossing between levels with $m_s=0$ and $m_s=-1$
for $^3$He-NH as a function of magnetic field $B$.} \label{figac3}
\end{figure}

As expected, the Zeeman effect is generally quite linear for He-NH
in the range of fields studied. This is because for $n=0$ the only
off-diagonal terms are those of $V(R,\theta)$ and $\hat H_{\rm
SS}$ that mix in excited $n$ levels, and the spacing between the
$n=0$ and $n=1$ levels for NH is around 32.6 cm$^{-1}$.
Nevertheless, there are avoided crossings where levels of the same
$M_{\cal J}$ and parity but different $m_s$ cross. Expanded views
of the energy level diagrams in the region of avoided crossings
are shown for $^4$He in Fig.\ \ref{figac4} and for $^3$He in Fig.\
\ref{figac3}. It may be seen that in this system the crossings are
very tightly avoided, with spacings at the crossing points of
$\Delta E = 5.2\times 10^{-4}$ cm$^{-1}$ and $\Delta E=2.0 \times
10^{-3}$ cm$^{-1}$ for $^4$He and $\Delta E=1.5 \times 10^{-3}$
cm$^{-1}$ for $^3$He.

The reason that the crossings are so tightly avoided in this case
is that there are {\it no} direct off-diagonal matrix elements
between $n=0$ basis functions with different values of $m_s$. The
dominant coupling is a second-order one of the form
\begin{equation}
\frac{\langle 001m_s| \hat H_{\rm SS}|2m_n1m_s'\rangle \langle
2m_n1m_s'|V_2|001m_s'\rangle}{E_{n=2}-E_{n=0}},
\end{equation}
where basis functions are represented $|nm_nsm_s\rangle$ and
$m_n=m_s-m_s'$. Such crossings will thus be more strongly avoided
in systems with stronger anisotropy or smaller monomer rotational
constant.

If matrix elements of $\hat H_{\rm SS}$ off-diagonal in $n$ are
omitted, the avoided crossings still exist but are about a factor
of 5 tighter. Under these circumstances they are caused by a
third-order mechanism in which potential couplings mix in basis
functions with $n>0$ (but the same values of $m_s$ and $m_s'$) and
these are connected by $\hat H_{\rm SS}$.

It should be noted that for levels with $n>0$ there are {\it
direct} matrix elements of $\hat H_{\rm SS}$ and $\hat H_{\rm NS}$
that connect levels with different $m_s$ but the same $m_j$. This
will produce more strongly avoided crossings where the separation
is simply proportional to $\lambda_{\rm SS}$ (or $\gamma$ for
$^2\Sigma$ monomers). This will be particularly important for
$^{16}$O$_2$, which has an $n=1$ ground state.

The boundary conditions applied by the BOUND program are only
correct for true bound states, below the lowest dissociation
threshold. Above a threshold, the boundary conditions produce
artificially quantized states. These are of two types: states that
are predominantly in an {\it open} channel, which have no physical
significance; and states that are predominantly in a {\it closed}
channel, which correspond closely to quasibound states of the real
system but with only an approximate open-channel (dissociative)
component. The two types are very easy to tell apart: the
open-channel states are closely parallel to a lower threshold, and
inspection of the wavefunction confirms their parentage. The
open-channel states have been omitted from Figs.\ 1 and 2. The
closed-channel states are shown as dashed lines above the lowest
threshold. They allow us to estimate very precisely where a bound
or quasibound state crosses a threshold, and thus where a
zero-energy resonance is expected in the scattering. This
information will be used in section \ref{scatsec} below.

\section{Methods for scattering calculations}

The coupled equations needed for scattering calculations are
identical to those for bound states. The only differences are that
the energy $E$ is above one or more thresholds $E_i$ and that
scattering boundary conditions are applied at long range. In the
present work we solve the coupled equations for scattering using
the MOLSCAT package \cite{molscat:v14}. We have generalised
MOLSCAT in the same way as BOUND to handle basis sets in which the
monomer Hamiltonian is nondiagonal.

\subsection{Boundary conditions}

The usual procedure for obtaining the scattering matrix $S$ from
the log-derivative matrix $Y$ has been described by Johnson
\cite{Johnson:1973}. Each channel is either open, $E\ge E_i$ or
closed, $E<E_i$. If matching takes place at a finite distance
$R_{\rm max}$ where the wavefunction in the closed channels
$(E<E_i)$ has not decayed to zero, it is necessary to take account
of the closed channels. The asymptotic form of the wavefunction is
\begin{equation}
\chi(R) = J(R) + N(R) K, \label{sboundary}
\end{equation}
where $J(R)$ and $N(R)$ are diagonal matrices made up
Riccati-Bessel functions for open channels and modified spherical
Bessel functions for closed channels. If there are $N$ channels
and $N_{\rm open}$ open channels, the $N\times N$ log-derivative
matrix is converted into an $N\times N$ real symmetric $K$ matrix.
The $S$ matrix is then obtained from the open-open submatrix,
$K_{\rm oo}$, using the relationship
\begin{equation}
S=(I + {\rm i} K_{\rm oo})^{-1} (I - {\rm i} K_{\rm oo}),
\end{equation}
where $I$ is an $N_{\rm open} \times N_{\rm open}$ unit matrix.
However, the boundary conditions (\ref{sboundary}) are appropriate
only in a basis set in which both $\hat L^2$ and the asymptotic
Hamiltonian are diagonal. In our generalised version of MOLSCAT,
the log-derivative matrix $Y$ is propagated in the primitive basis
set (which is nondiagonal) and then transformed at $R_{\rm max}$
into a basis set that diagonalises $\hat H_{\rm mon} + \hat H_{\rm
Z}$.

For the simple case of He-NH, the eigenvalues of $\hat H_{\rm mon}
+ \hat H_{\rm Z}$ are nondegenerate except at non-zero field. The
transformation that diagonalises $\hat H_{\rm mon} + \hat H_{\rm
Z}$ is thus unique. However, in more complicated cases with two
structured collision partners it will be necessary to transform to
a basis set that diagonalises $\hat H_{\rm mon,1} + \hat H_{\rm
Z,1}$ and $\hat H_{\rm mon,2} + \hat H_{\rm Z,2}$ separately.
Additional degeneracies arise at zero field, and resolving them
may require diagonalisation of another operator such as $\hat
j_Z$.

\subsection{Cross sections}

The cross section for a transition $i\rightarrow f$ from initial
state $i$ to final state $f$ is obtained from the square of the
corresponding $T$ matrix element,
\begin{equation}
\sigma_{if} = \frac{\pi}{k^2} |T_{if}|^2,
\end{equation}
where $k$ is the incoming wave vector, $k^2=2\mu (E-E_i)/\hbar^2$,
and $T_{if}=\delta_{if}-S_{if}$. In general it is necessary to sum
over all channels corresponding to the monomer levels of interest
and over $S$ matrices obtained for different values of $M_{\cal
J}$ and parity.

\subsection{Scattering lengths}

In the ultracold regime, scattering properties are often described
in terms of a complex scattering length $a=\alpha-{\rm i}\beta$
\cite{Balakrishnan:scat-len:1997, Bohn:1997}. The diagonal
S-matrix element in the incoming channel $0$ may be written in
terms of a complex phase shift $\delta$ \cite{Mott:p380:1965},
\begin{equation}
S_{00}=\exp(2{\rm i}\delta) \label{eqsd}
\end{equation}
and the complex scattering length is defined by
\begin{equation}
a= \frac{-\tan\delta}{k}. \label{eqad}
\end{equation}
Equivalently,
\begin{equation}
S_{00} = \frac{1-{\rm i}ka}{1+{\rm i}ka}.
\end{equation}
The scattering length becomes constant at limitingly low energy.
The elastic and total inelastic cross sections are
\cite{Cvitas:li3:2006}
\begin{equation}
\sigma_{\rm el} = \frac{4\pi|a|^2}{1+k^2|a|^2+2k\beta}
\end{equation}
and
\begin{equation}
\sigma_{\rm inel} = \frac{4\pi\beta}{k(1+k^2|a|^2+2k\beta)}.
\end{equation}

The scattering length is often given as
\begin{equation}
a=\lim_{k\rightarrow0} \frac{1}{2{\rm i}} \frac{T_{00}}{k}.
\label{eqat}
\end{equation}
However, this relies on a Taylor series expansion of $\exp(2{\rm
i}\delta)$ that is valid only when $\delta\ll 1$. Across an
elastic scattering resonance, $\delta$ changes by $\pi$ even at
limitingly low energy, so Eq.\ \ref{eqat} is inappropriate. In the
present work we obtain scattering lengths numerically either by
converting the low-energy S-matrix elements to complex phases
using Eq.\ \ref{eqsd} and then the definition (\ref{eqad}) or by
using the equivalent identity
\begin{equation}
a = \frac{1}{{\rm i}k} \left(\frac{1-S_{00}}{1+S_{00}}\right).
\end{equation}

\subsection{Resonant behavior}

If there is only one open channel, then the phase shift $\delta$
is real and its behavior is sufficient to characterize a
resonance. It follows a Breit-Wigner form as a function of energy,
\begin{equation}
\delta(E) = \delta_{\rm bg} + \tan^{-1}
\left[\frac{\Gamma_E}{2(E_{\rm res}-E)}\right], \label{eqbw}
\end{equation}
where $\delta_{\rm bg}$ is a slowly varying background term,
$E_{\rm res}$ is the resonance position and $\Gamma_E$ is its
width (in energy space). This corresponds to the $S$ matrix
element describing a circle of radius 1 in the complex plane. In
general the parameters $\delta_{\rm bg}$, $E_{\rm res}$ and
$\Gamma_E$ are slow functions of energy, but this is neglected in
the present work apart from threshold behaviour.

The resonance position $E_{\rm res}$ and the threshold energy
$E_{\rm thresh}$ are both functions of magnetic field,
\begin{equation}
\frac{dE_{\rm res}}{dB}=\mu_{\rm res} \quad \hbox{and} \quad
\frac{dE_{\rm thresh}}{dB}=\mu_{\rm thresh}.
\end{equation}
We define $B_{\rm res}(E)$ as the field at which $E_{\rm res}=E$.
For low kinetic energies, the width $\Gamma_E$ also depends on the
field through its threshold dependence \cite{Timmermans:1999},
\begin{equation}
\Gamma_E(E_{\rm kin})=2k\gamma_E,
\end{equation}
with constant reduced width $\gamma_E$. As a function of magnetic
field at constant kinetic energy, the phase shift thus follows a
form similar to Eq.\ \ref{eqbw},
\begin{equation}
\delta(B) = \delta_{\rm bg} + \tan^{-1}
\left[\frac{\Gamma_B(E_{\rm kin})}{2(B_{\rm res}(E)-B)}\right].
\label{eqbwb}
\end{equation}
The width $\Gamma_B(E_{\rm kin})$ is a signed quantity that is
negative if the bound state tunes upwards through the energy of
interest and positive if it tunes downwards,
\begin{equation}
\Gamma_B(E_{\rm kin}) = \frac{\Gamma_E(E_{\rm kin})}{\mu_{\rm
thresh} - \mu_{\rm res}}.
\end{equation}
The background phase shift $\delta_{\rm bg}$ goes to zero as
$k\rightarrow0$ according to Eq.\ \ref{eqad} (with constant finite
$a_{\rm bg}$), but the resonant term still exists. The scattering
length passes through a pole when
$\delta=\left(n+\frac{1}{2}\right)\pi$. The scattering length
follows the formula commonly used in atomic scattering
\cite{Moerdijk:1995},
\begin{equation}
a(B) = a_{\rm bg} \left[ 1 - \frac{\Delta_B}{B - B_{\rm res}(E)}
\right], \label{eqares}
\end{equation}
where $\Delta_B=-\Gamma_B/2ka_{\rm bg}=-\gamma_B/a_{\rm bg}$ and
is independent of $k$ near threshold.

When there are several open channels, $\delta$ is in general
complex. The quantity that then follows the Breit-Wigner form
(\ref{eqbw}) or (\ref{eqbwb}) is the S-matrix eigenphase sum
\cite{Ashton:1983}, which is the sum of phases of the {\it
eigenvalues} of the $S$ matrix. The eigenphase sum is real,
because the $S$ matrix is unitary, so that all its eigenvalues
have modulus 1. In practice the eigenphase sum is most
conveniently calculated by diagonalising the real symmetric matrix
$K_{\rm oo}$ and summing the inverse tangents of its eigenvalues.

If there is more than one open channel, the individual $S$ matrix
elements still describe circles in the complex plane,
\begin{equation}
S_{ii'} = S_{{\rm bg,}ii'} - \frac{{\rm i} g_{Ei} g_{Ei'}}{E -
E_{\rm res} + {\rm i}\Gamma_E/2},
\end{equation}
where $g_{Ei}$ is complex. The {\it partial width} for channel $i$
is $\Gamma_{Ei}=|g_{Ei}|^2$ and the radius of the circle in
$S_{ii'}$ is $|g_{Ei} g_{Ei'}|/\Gamma_E$. The analogous expression
as a function of magnetic field at constant kinetic energy is
\begin{equation}
S_{ii'} = S_{{\rm bg,}ii'} - \frac{{\rm i} g_{Bi} g_{Bi'}}{B -
B_{\rm res} + {\rm i}\Gamma_B/2}, \label{eqsii}
\end{equation}
where the energy-dependence of $B_{\rm res}$ and $\Gamma_B$ has
been omitted to simplify notation. If $\Gamma_{E0}<\Gamma_E$ (or
equivalently $|\Gamma_{B0}| < |\Gamma_B|$), the scattering length
does {\it not} pass through a pole. In the low-energy threshold
regime, $\Gamma_{B0}$ is proportional to $k$ and we may define an
energy-independent {\it reduced width} $\gamma_{B0}$,
\begin{equation}
\Gamma_{B0} = 2 \gamma_{B0} k.
\end{equation}
However, $\Gamma_B$ has inelastic contributions $\Gamma_B^{\rm
inel}$ that are essentially energy-independent. Hutson
\cite{Hutson:res:2006} has defined a {\it resonant scattering
length},
\begin{equation}
a_{\rm res} = \frac{2\gamma_{B0}}{\Gamma_B^{\rm inel}},
\end{equation}
that characterizes the strength of the resonant contribution to
the scattering at low energy. If $ka_{\rm res} \ll 1$, $S_{00}$
describes a circle of radius $2ka_{\rm res}$ in the complex plane
as a resonance is tuned through threshold and the real part of the
scattering length oscillates by $\pm a_{\rm res}/2$.

As will be seen below, for He + NH the inelastic scattering
strongly suppresses the pole in the scattering length and the
resonant oscillation in the scattering length is of quite small
amplitude. The corresponding oscillations in the elastic and
inelastic cross sections are also relatively weak.

\section{Results of scattering calculations}
\label{scatsec}

We have carried out scattering calculations on $^4$He-NH and
$^3$He-NH using the potential energy surface of Krems {\em et
al.}\ \cite{Krems:henh:2003}. These calculations used a reduced
basis set of functions with $n \le 4$ and $L \le 5$ to allow
comparison with previous studies. The coupled equations were
propagated outwards from 1.7 \AA\ to 120.0 \AA\ using Johnson's
log-derivative algorithm \cite{Johnson:1973} with a sector size of
0.025 \AA.

We have verified that the new code gives identical scattering
results for He-NH with the coupled and uncoupled basis sets. This
confirms the correctness of the coding for the basis sets and the
extraction of $S$ matrices. Our program also gives identical
results to refs.\ \onlinecite{Krems:henh:2003} and
\onlinecite{Cybulski:2005} for scattering in a magnetic field.

It is of great interest to investigate the effect of zero-energy
Feshbach resonances on low-energy molecular scattering as a
function of magnetic field. However, a major problem is that the
resonances can be very narrow, and locating the fields at which
they occur is difficult and time-consuming. For example, in the
He-NH problem we expect the coupling between bound and continuum
states with different $m_j$ to be comparable to or smaller than
that between bound states of different $m_j$. To a first
approximation, the latter is the energy separation of the
bound-state avoided crossings, which is around $10^{-3}$
cm$^{-1}$. Since the energy of a state with $m_s=\pm1$ tunes by
about $10^{-4}$ cm$^{-1}$/G, we expect the Feshbach resonances to
be less (perhaps much less) than 10 G wide.

Fortunately, the bound-state capability described above provides a
solution to this problem: we can extrapolate the bound-state
energies to calculate the field at which they cross a threshold
and then scan across a small range of fields in the vicinity. As
an example, we searched for Feshbach resonances in $^3$He-NH. It
may be seen in Fig.\ \ref{figbs3} that there is a $^3$He-NH bound
state with $m_s=+1$ that crosses the $m_s=-1$ threshold at about
7200 G and the $m_s=0$ threshold at about 14300 G. Careful
extrapolation of the bound-state energies with the basis set used
for the scattering calculations gives more precise field estimates
of 7168.750 G (for $M_{\cal J}=-1$) and 14340.36 G (for $M_{\cal
J}=0$) respectively.

\begin{figure}[tb]
\includegraphics[width=80mm]{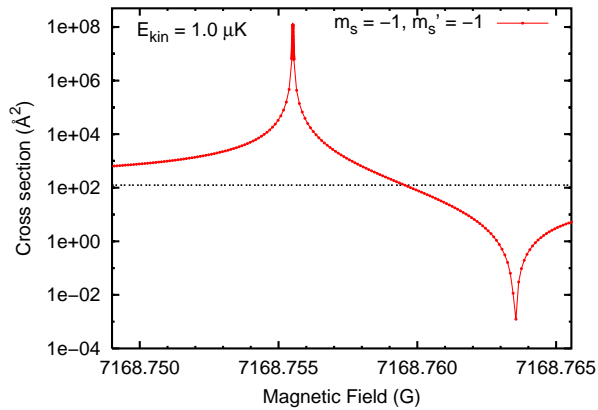}
\caption{Elastic ($m_s= -1\rightarrow -1$) cross section for
$^3$He-NH collisions in the vicinity of an elastic Feshbach
resonance at a kinetic energy of $10^{-6}$~K.} \label{figxs1}
\end{figure}

\begin{figure}[tb]
\includegraphics[width=80mm]{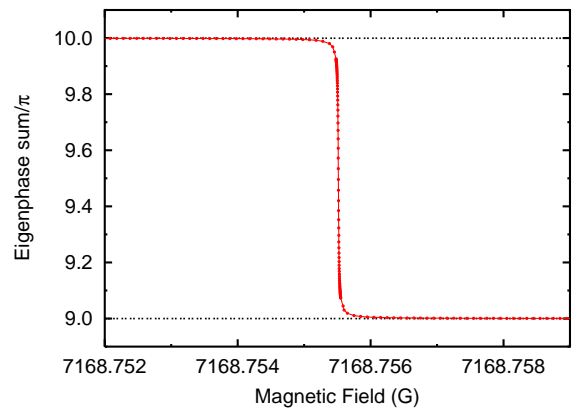}
\caption{Phase shift for elastic $^3$He-NH collisions in the
vicinity of an elastic Feshbach resonance at a kinetic energy of
$10^{-6}$~K.} \label{figes1}
\end{figure}

Care must be taken to use bound-state calculations for the correct
values of $M_{\cal J}$ and parity. In the present case, we want
s-wave resonances, with $L=0$ in the incoming channel. This
requires $M_{\cal J} = m_j$ and even parity. Bound-state
calculations with different values of $M_{\cal J}$ produce
energies that cross threshold at different values of the field. In
He-NH, which is a very weakly coupled system with almost linear
Zeeman effects, the fields are only slightly different; for
example, the crossing with the $m_s = 0$ threshold occurs at
14345.40 G for $M_{\cal J}=-1$. Nevertheless, the difference can
easily be enough to miss the resonance, and in more strongly
coupled systems will be crucial.

\begin{figure}[tb]
\includegraphics[width=80mm]{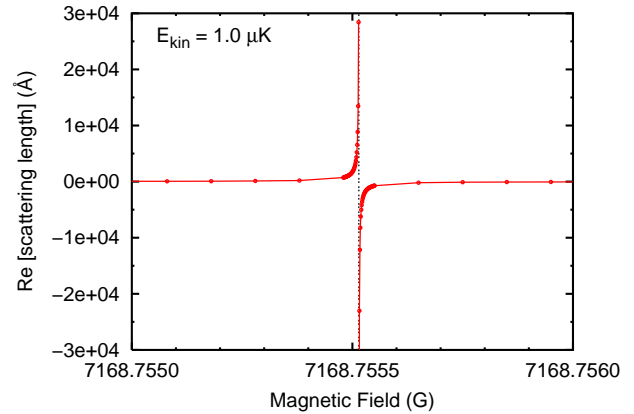}
\caption{Scattering length for elastic $^3$He-NH collisions in the
vicinity of an elastic Feshbach resonance, from calculations at a
kinetic energy of $10^{-6}$~K.} \label{figsl1}
\end{figure}

\begin{figure}[tb]
\includegraphics[width=80mm]{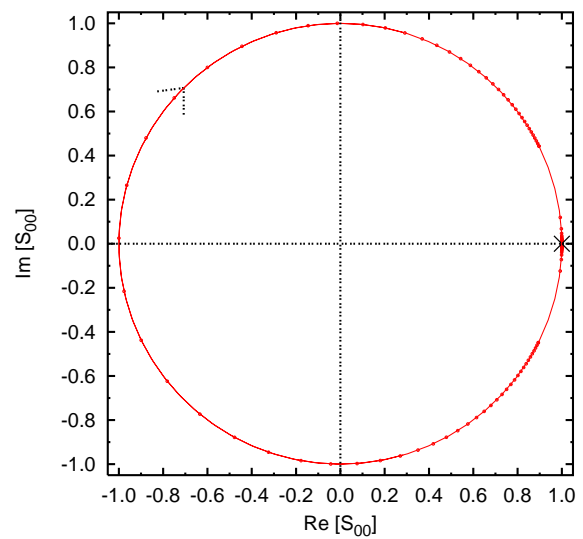}
\caption{The circle of radius 1 in the complex plane described by
the $L=0$ diagonal $S$ matrix element for elastic scattering in
the vicinity of an elastic Feshbach resonance at a kinetic energy
of $10^{-6}$~K. The cross shows the value far from resonance.}
\label{figsm1}
\end{figure}

For $^3$He colliding with NH ($m_s=-1$) near 7169 G, only elastic
scattering can occur. For $M_{\cal J}=-1$ and even parity, the
present basis set gives 3 open channels with $L=0$, 2 and 4.
Scattering into the $L>0$ channels is strongly suppressed by the
centrifugal barriers. The elastic cross section is shown as a
function of field in Fig.\ \ref{figxs1} and the corresponding
eigenphase sum and scattering length are shown in Figs.\
\ref{figes1} and \ref{figsl1}. The peak in the cross section for
kinetic energy $E_{\rm kin}=10^{-6}$~K is close to the value of
$4\pi/k^2=1.2\times10^8$~\AA$^2$ characteristic of a pole in the
scattering length. The peak shows an asymmetric Fano lineshape
\cite{Fano:1961}, with interference between a background term and
a resonant term that interfere constructively on the low-field
side of the resonance and destructively on the high-field side.
The diagonal $S$ matrix element for $L=0$ describes a circle of
radius 1 in the complex plane as the field is ramped across the
resonance, as shown in Fig.\ \ref{figsm1}. Fitting the eigenphase
sum to Eq.\ \ref{eqbwb} gives $\delta_{\rm bg} < 0.001$, $B_{\rm
res}=7168.7555$~G and $\Gamma_B = -1.65 \times 10^{-5}$~G, while
fitting the scattering length to Eq.\ \ref{eqares} gives $a_{\rm
bg}=3.19$~\AA, $B_{\rm res}=7168.7555$~G and $\Delta_B=8.04\times
10^{-3}$ G. The widths $\Gamma_B$ and $\Delta_B$ are related by
$\Gamma_B/2=-ka_{\rm bg}\Delta_B$. It may be seen that the
bound-state calculation does indeed give a very precise estimate
of the position of the zero-energy resonance.
\begin{figure}[tb]
\includegraphics[width=80mm]{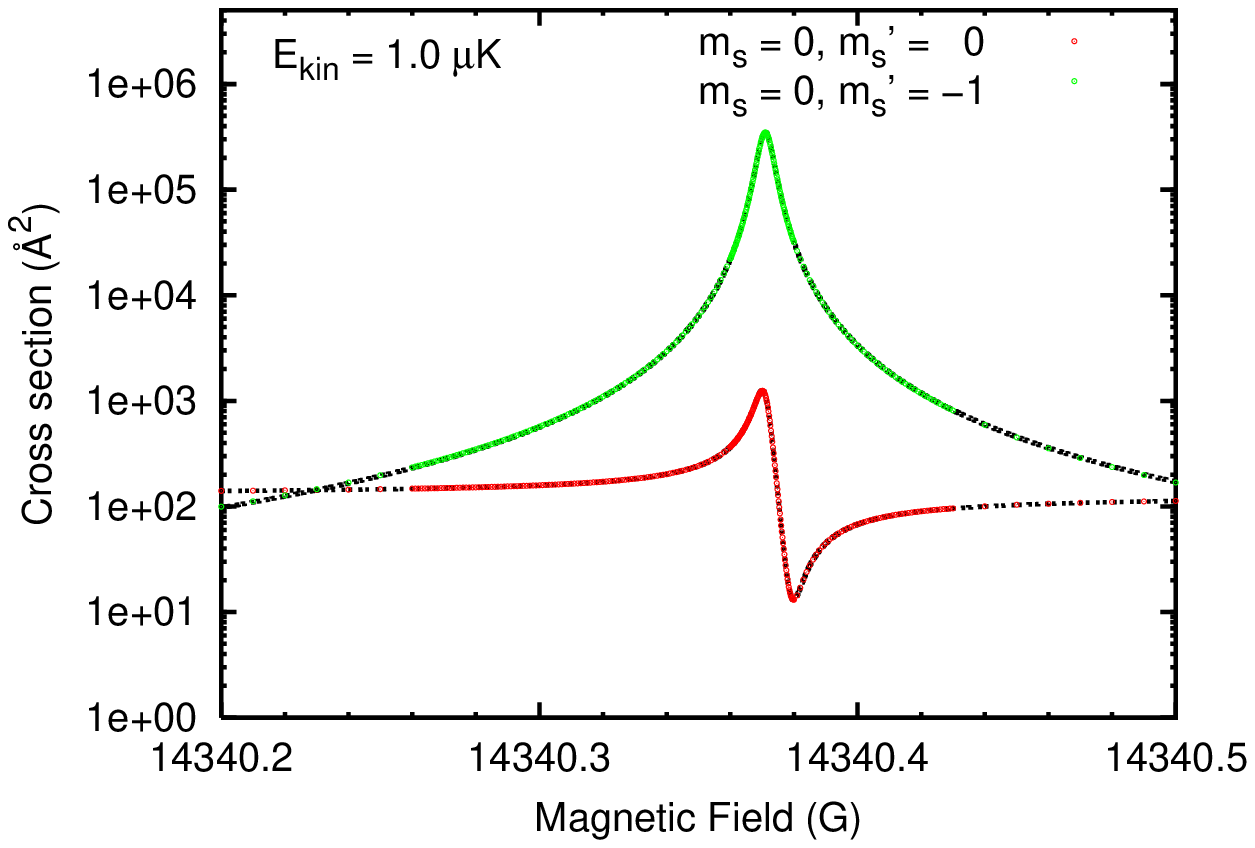}
\includegraphics[width=80mm]{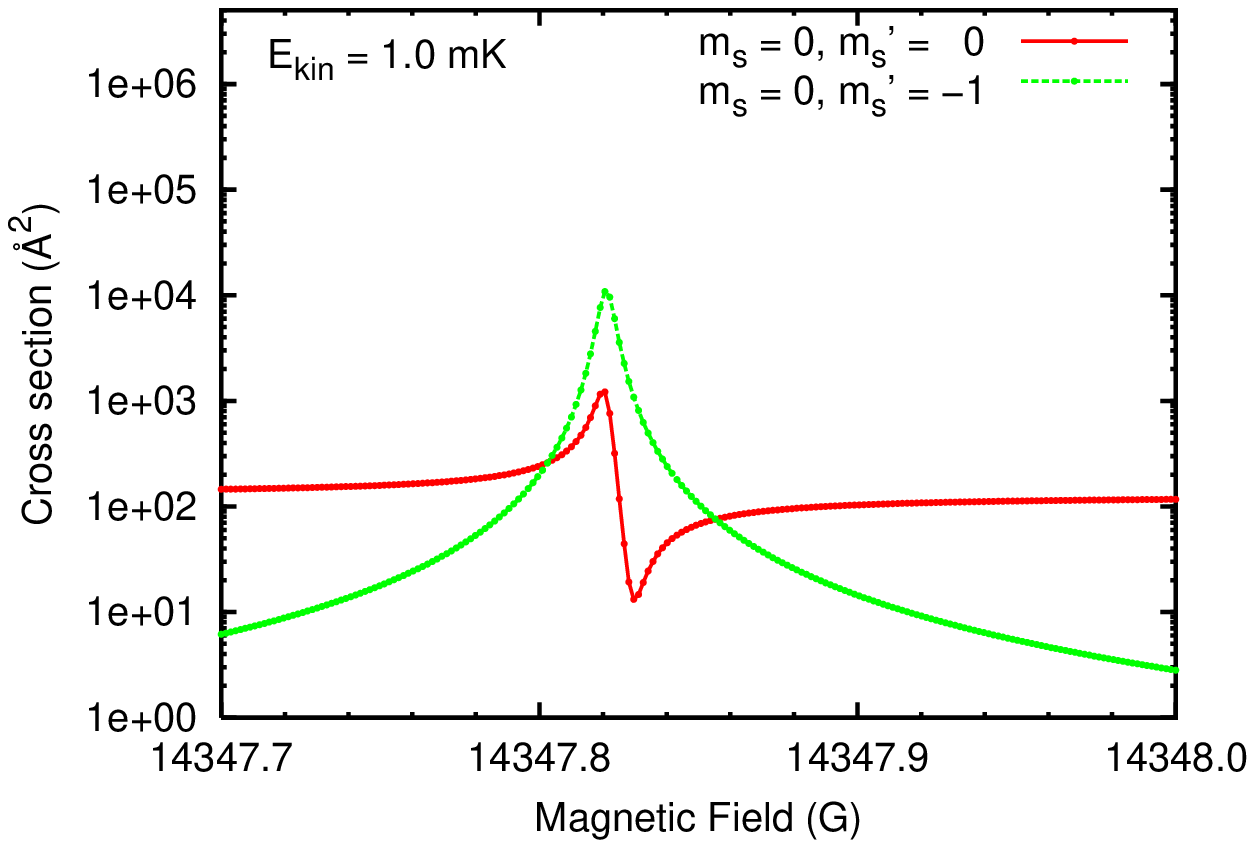}
\caption{Elastic ($m_s= 0\rightarrow 0$) and inelastic ($m_s = 0
\rightarrow -1)$ cross sections for $^3$He-NH collisions in the
vicinity of an inelastic Feshbach resonance at kinetic energies of
$10^{-6}$~K (upper panel) and $10^{-3}$~K (lower panel). The lines
show the results of Eqs.\ \ref{eqsigel} and \ref{eqsiginel}.}
\label{figxs2}
\end{figure}

For $^3$He colliding with NH ($m_s=0$) near 14300 G, both elastic
and inelastic scattering are possible. For $M_{\cal J}=0$ and even
parity, our basis set gives 3 elastic channels ($m_s=0$ with
$L=0$, 2 and 4) and 2 inelastic channels ($m_s=-1$ with $L=2$ and
4). However, the elastic channels with $L>0$ make no significant
contributions at ultralow energies. Fig.\ \ref{figxs2} shows scans
of the elastic ($m_s= 0\rightarrow 0$) and total inelastic ($m_s =
0 \rightarrow -1)$ cross sections for kinetic energies of
$10^{-6}$~K and $10^{-3}$~K. Once again the bound-state
calculation gives a very precise estimate of the position of the
zero-energy resonance. At $10^{-3}$~K the resonance is shifted
slightly because a different field is needed to bring the bound
state into resonance with the larger total energy. Apart from the
shift, however, the cross sections behave as expected from the
Wigner threshold laws: the elastic cross section is almost
unchanged and the inelastic cross section scales with $k^{-1}$.

\begin{figure}[tb]
\includegraphics[width=80mm]{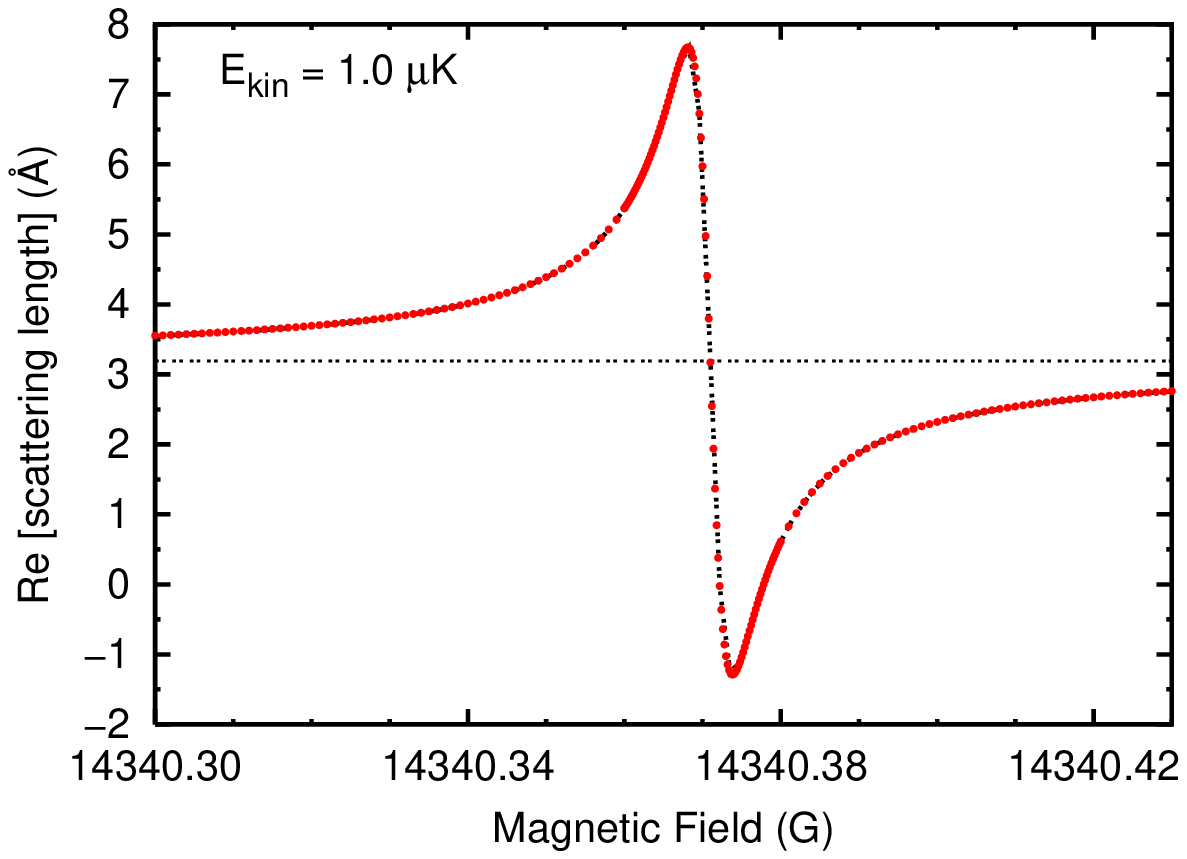}
\includegraphics[width=80mm]{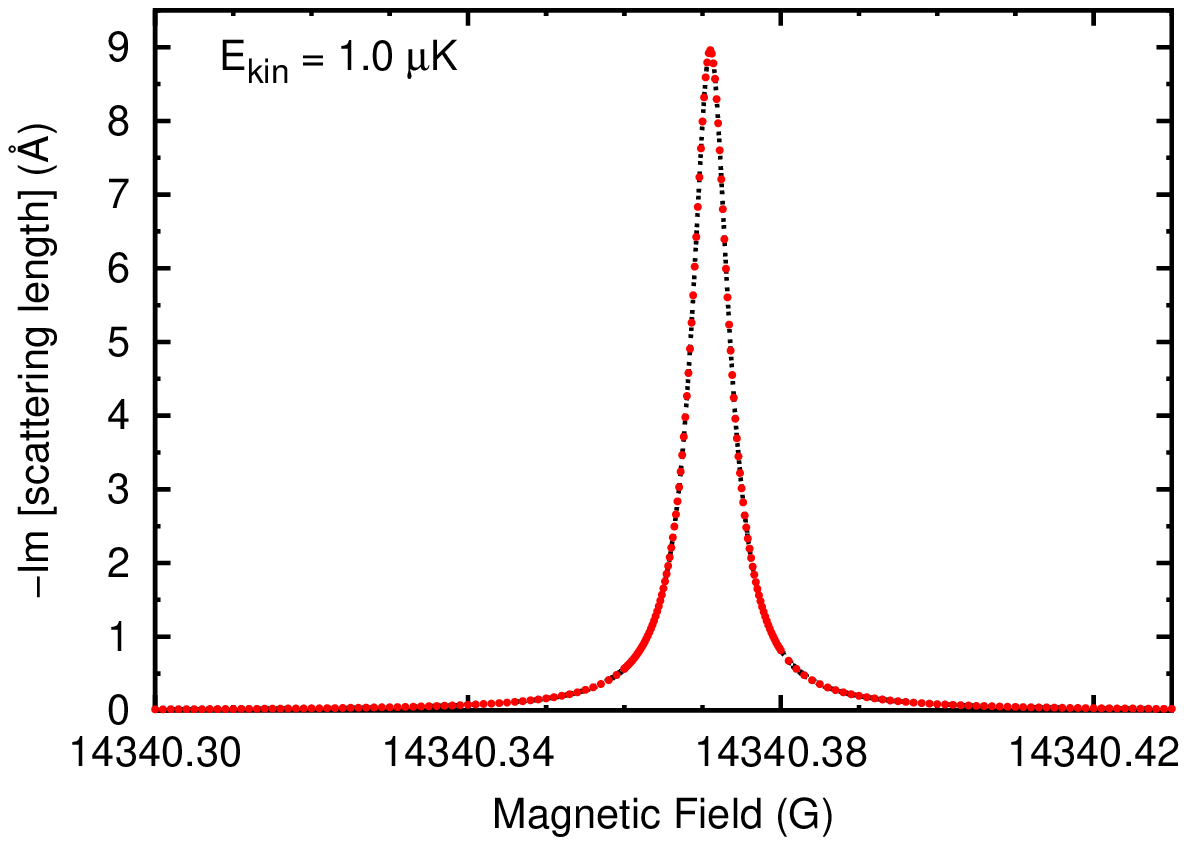}
\caption{Real and imaginary parts of the scattering length for
$^3$He-NH collisions in the vicinity of an inelastic Feshbach
resonance, from calculations at a kinetic energy of $10^{-6}$~K.
Upper panel: real part; lower panel: imaginary part. The lines
show the results of Eq.\ \ref{eqaares}.} \label{figsl2}
\end{figure}

\begin{figure}[tb]
\includegraphics[width=80mm]{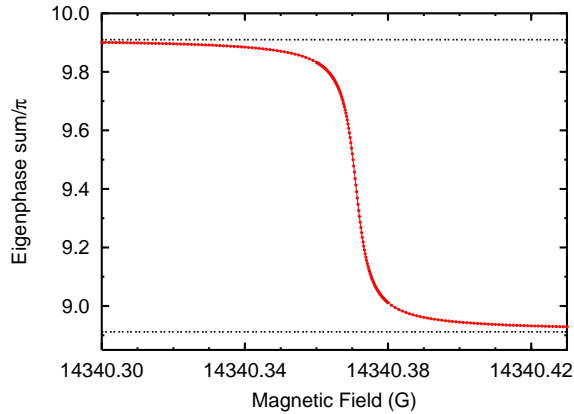}
\caption{Eigenphase sum for inelastic $^3$He-NH collisions in the
vicinity of an inelastic Feshbach resonance at a kinetic energy of
$10^{-6}$~K.} \label{figes2}
\end{figure}

A notable feature of Fig.\ \ref{figxs2} is that the elastic cross
section does {\it not} peak at the very high value of $4\pi/k^2$
characteristic of a pole in the scattering length. The real and
imaginary parts of the scattering length are shown in Fig.\
\ref{figsl2}. Instead of rising to $\infty$, the scattering length
oscillates with a peak at less than $+8$~\AA. The eigenphase sum
shows a sharp drop through $\pi$ as shown in Fig.\ \ref{figes2},
and fitting to Eq.\ \ref{eqbwb} gives $B_{\rm res}=14340.371$~G
and $\Gamma_B=-5.72\times 10^{-3}$~G. However, the phase change is
distributed between several diagonal matrix elements.

\begin{figure}[tb]
\includegraphics[width=79mm]{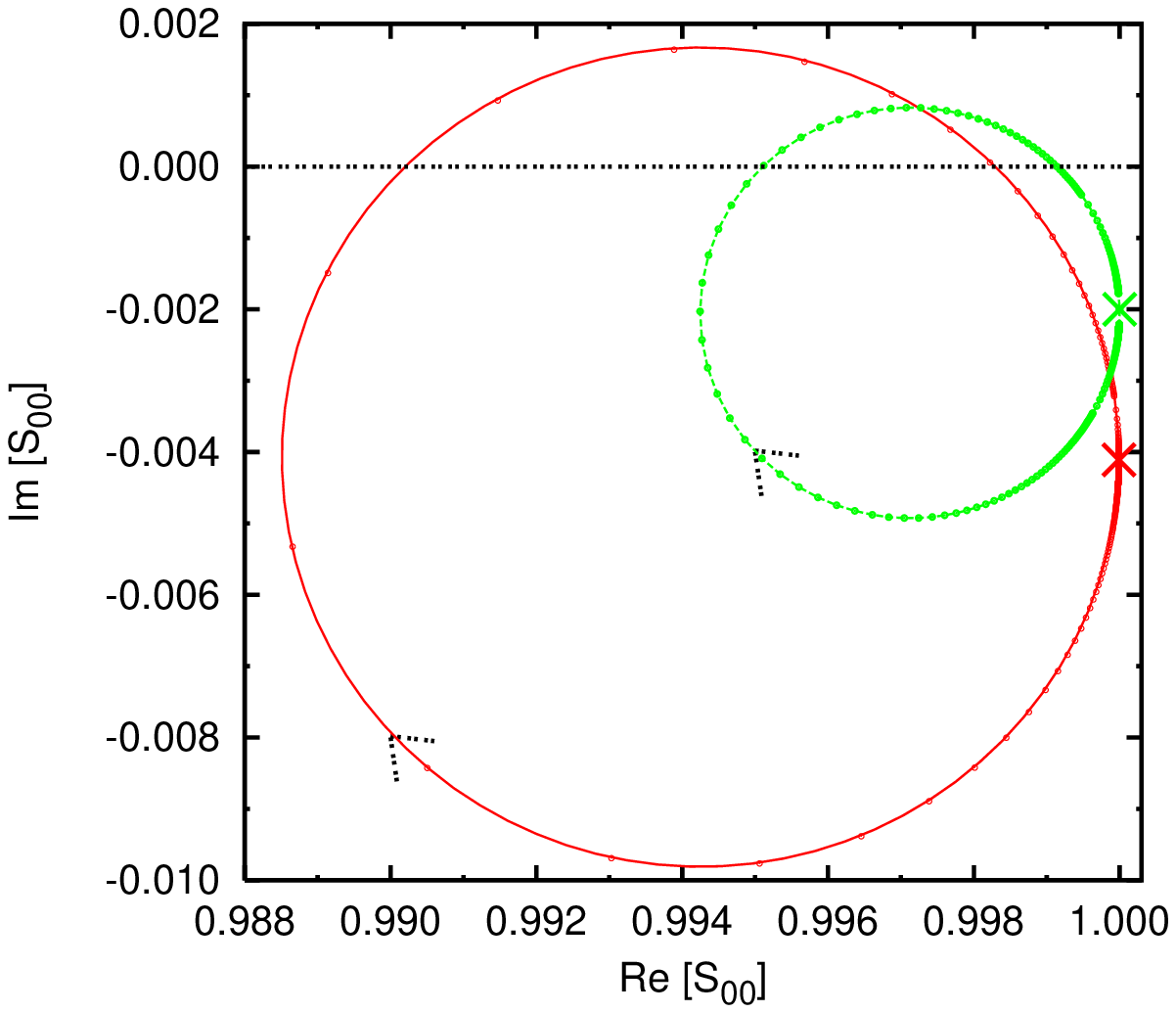}
\includegraphics[width=79mm]{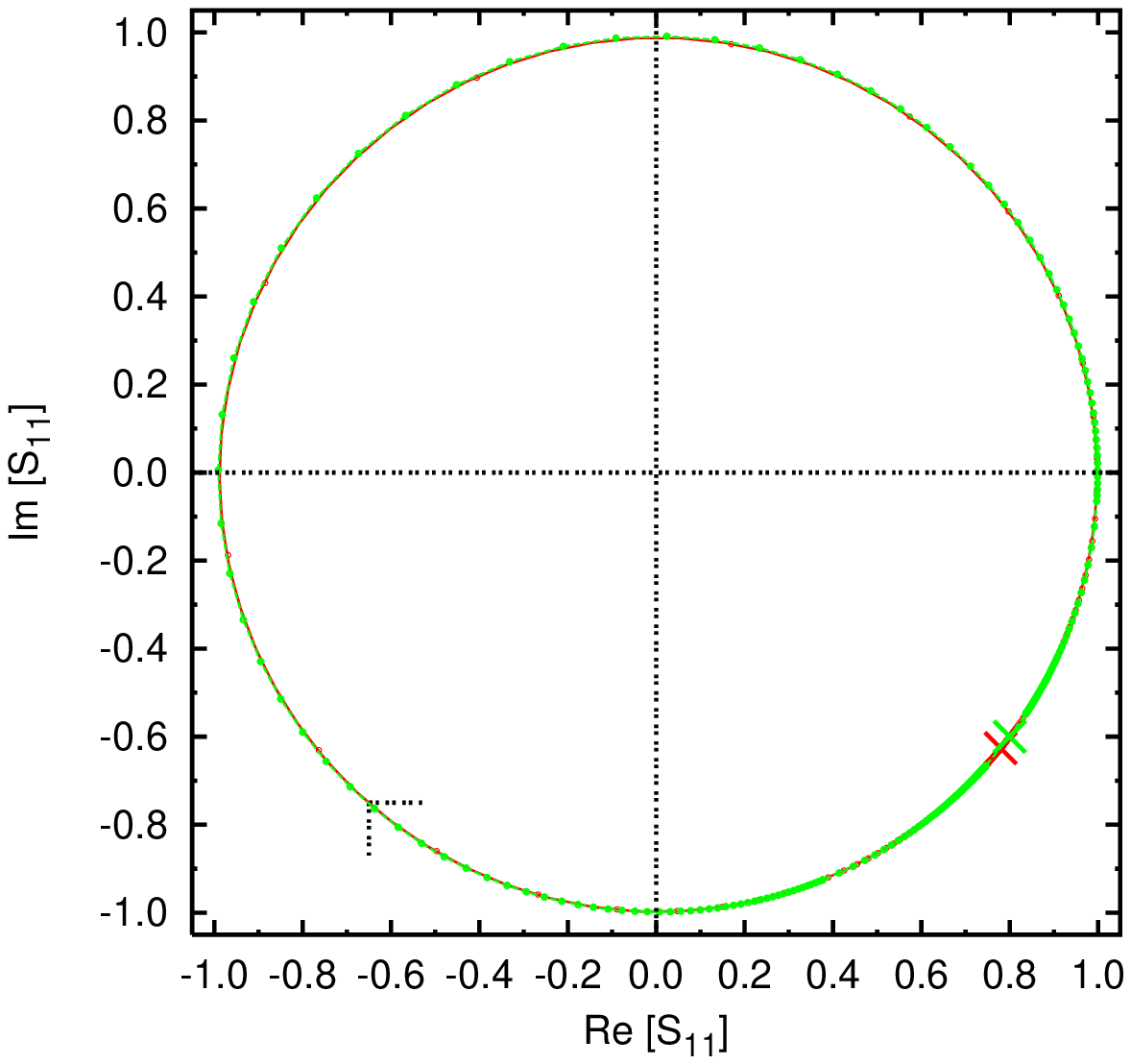}
\includegraphics[width=79mm]{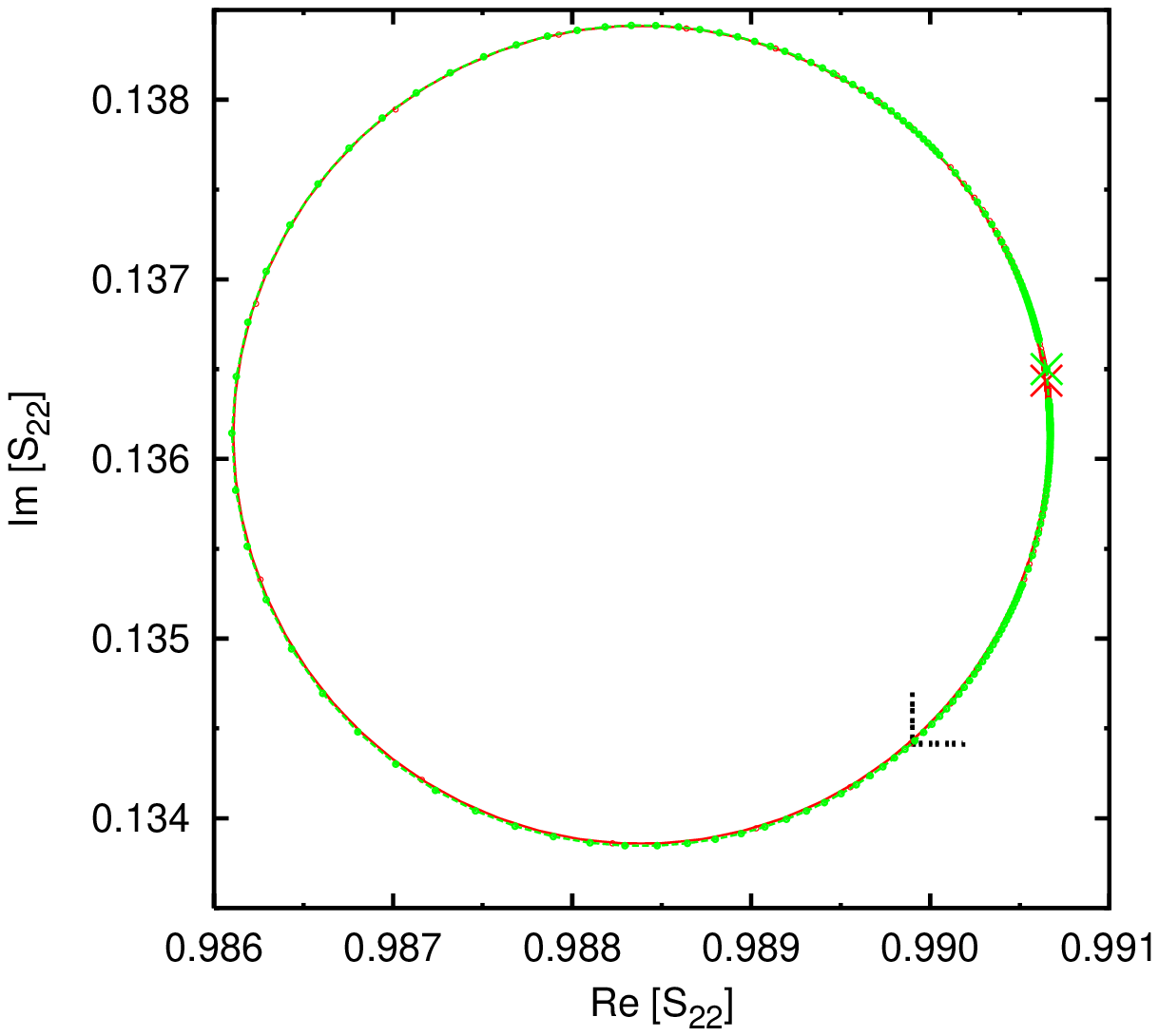}
\caption{The circles described by diagonal $S$ matrix elements in
the presence of inelastic scattering at $E_{\rm kin}=10^{-6}$~K
(green) and $4 \times 10^{-6}$~K (red). Upper panel: incoming wave
(channel 0: $m_s=0$, $L=0$); center panel: (channel 1: $m_s=-1$,
$L=2$); lower panel: (channel 2: $m_s=-1$, $L=4$). The crosses
show values far from resonance.} \label{figsm2el}
\end{figure}

\begin{figure}[tb]
\includegraphics[width=79mm]{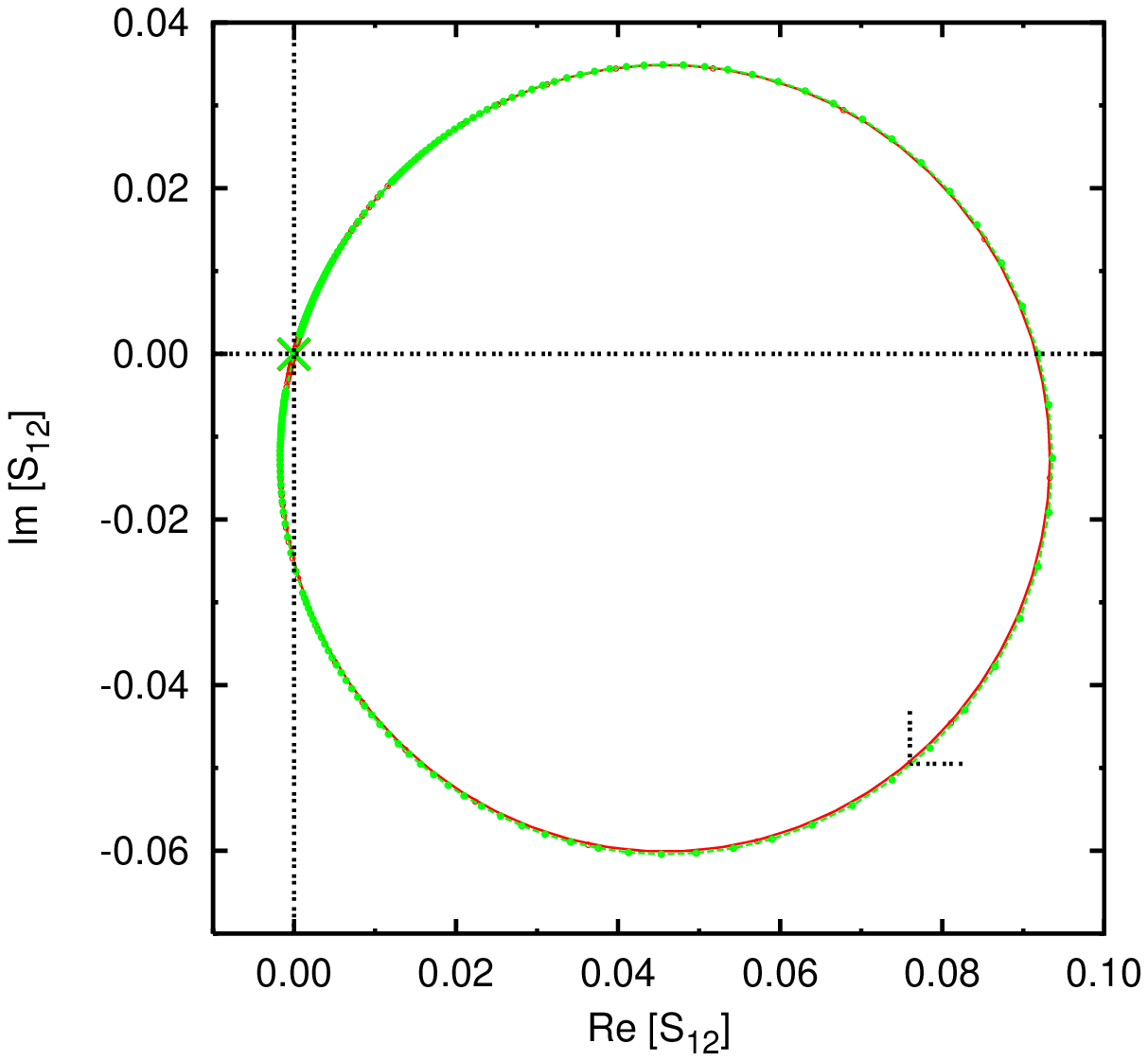}
\includegraphics[width=79mm]{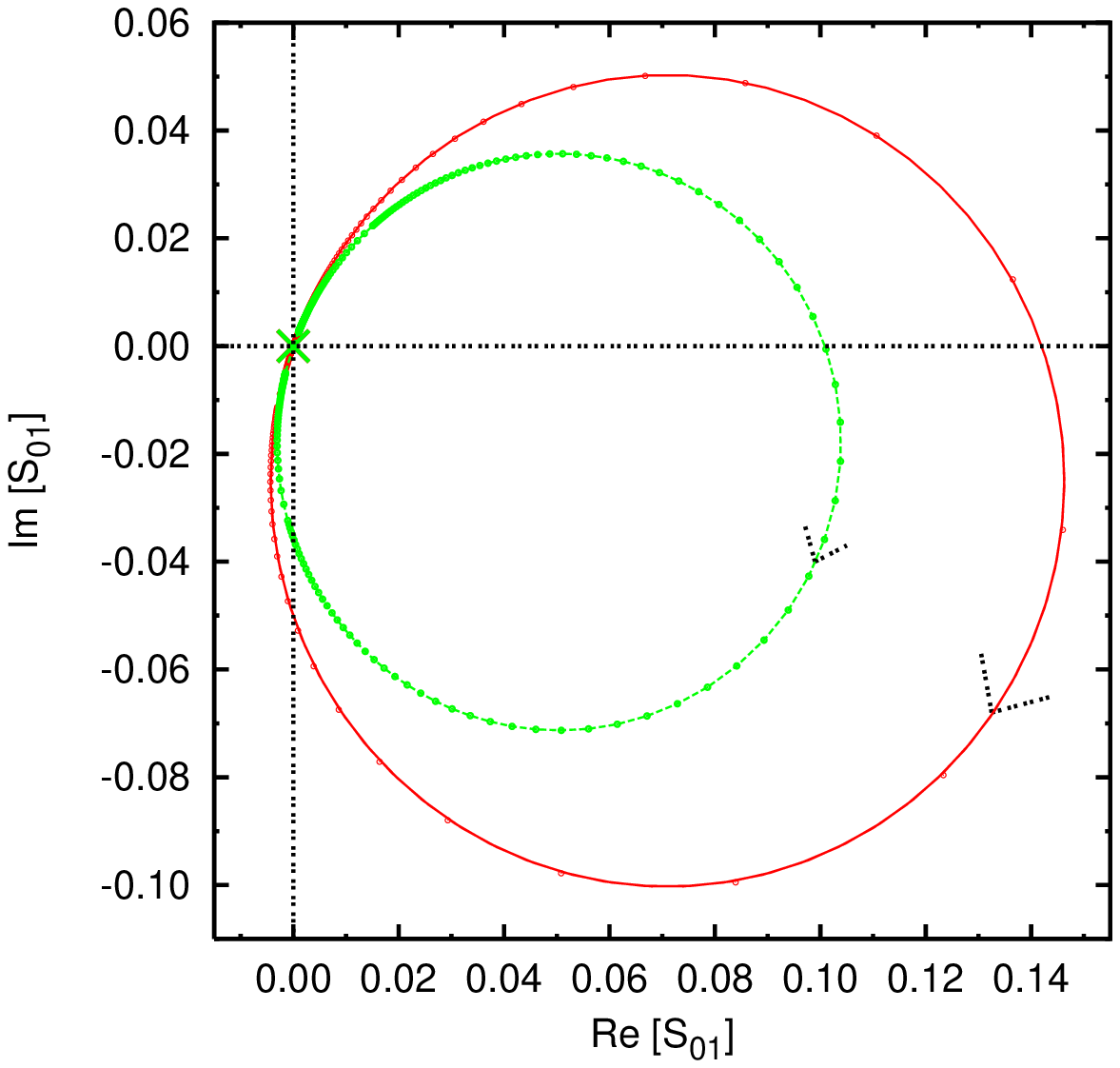}
\includegraphics[width=79mm]{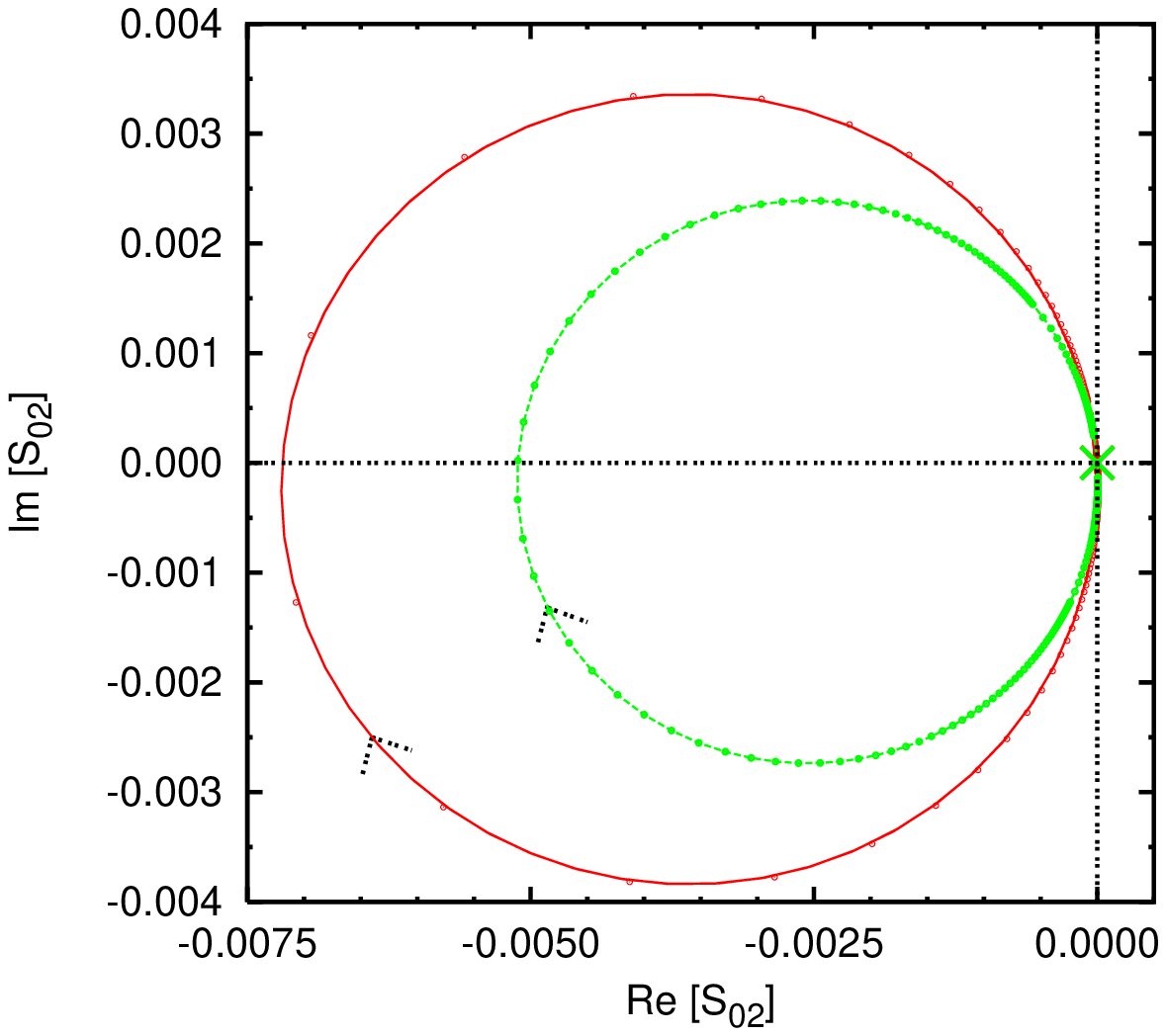}
\caption{The circles described by the off-diagonal $S$ matrix
elements for 3-channel scattering at $E_{\rm kin}=10^{-6}$~K
(green) and $4 \times 10^{-6}$~K (red). Channels are labeled as in
Fig.\ \ref{figsm2el}. The crosses show values far from resonance.}
\label{figsm2inel}
\end{figure}

The $S$ matrix elements are shown explicitly for two different
energies, $E_{\rm kin}= 10^{-6}$~K and $4 \times 10^{-6}$~K, in
Figs.\ \ref{figsm2el} and \ref{figsm2inel}. The corresponding
values of $k$ differ by a factor of 2. It may be seen in Fig.\
\ref{figsm2el} that the elastic $S$ matrix element for the
low-energy incoming channel describes a small circle with a radius
that depends linearly on $k$. For $E_{\rm kin}=10^{-6}$~K, the
radius is only about 0.003. The other diagonal elements describe
much larger circles that are almost independent of $k$, with
background (nonresonant) $S$ matrix elements (shown by crosses)
that are far from 1: these channels have substantial kinetic
energy so are not governed by the Wigner threshold laws. The
inelastic $S$ matrix elements are shown in Fig.\ \ref{figsm2inel},
and it may be seen that those for incoming channel 0 describe
circles with radius proportional to $k^{1/2}$.

Since $S_{00}$ describes a small circle in the complex plane, the
corresponding complex phase $\delta$ given by Eq.\ \ref{eqsd}
shows a relatively small oscillation that does not pass through
$\pm\pi/2$. There is thus an oscillation but no pole in the
scattering length $a$ (and no discontinuity in its sign). This
explains why the peak in the elastic scattering cross section is
much lower than the value of $4\pi/k^2$ expected when there is a
pole in $a$.

The behavior observed here may be quantified in terms of the
theory of Hutson \cite{Hutson:res:2006}. Fitting the individual
$S$ matrix elements at $E_{\rm kin}=10^{-6}$~K to Eq.\ \ref{eqsii}
gives a partial width $\Gamma_{B0} = -1.646 \times 10^{-5}$~G for
the incoming channel and $\Gamma_{B1} = -5.693 \times 10^{-3}$~G
and $\Gamma_{B2} = -1.308 \times 10^{-5}$~G for the two inelastic
channels. We have verified numerically that $\Gamma_{B0}$ is
proportional to $k$ and $\Gamma_{B1}$ and $\Gamma_{B2}$ are
independent of it. With $k=3.219 \times 10^{-4}$~\AA$^{-1}$, this
gives $\gamma_{B0} = 0.02557$ G~\AA\ and $a_{\rm res}=8.96$~\AA.
The peaks in the scattering lengths in Fig.\ \ref{figsl2} are at
$\alpha_{\rm max}=7.67$~\AA, $\alpha_{\rm min}=-1.28$~\AA, and
$\beta_{\rm max}=8.96$~\AA, which correspond to the theoretical
predictions \cite{Hutson:res:2006}, $\alpha^{\rm max}_{\rm
min}=a_{\rm bg}\pm a_{\rm res}/2$ and $\beta_{\rm max}=a_{\rm
res}$ with $a_{\rm bg}=3.19$~\AA.

The shapes of the peaks in scattering lengths and cross sections
also correspond very accurately to the theoretical profiles
\cite{Hutson:res:2006},
\begin{eqnarray}
a(B) &=& a_{\rm bg} + \frac{a_{\rm res}}{2(B-B_{\rm
res})/\Gamma_B^{\rm
inel}+{\rm i}}. \label{eqaares} \\
\sigma_{\rm el}(B)&\approx& 4\pi|a|^2 = 4\pi \left| a_{\rm bg} +
\frac{a_{\rm
res}}{2(B-B_{\rm res})/\Gamma_B^{\rm inel} + {\rm i}}\right|^2 \nonumber\\
\label{eqsigel} \\
\sigma_{\rm inel}^{\rm tot}(B)&\approx& \frac{4\pi\beta}{k} =
\frac{4\pi}{k} \frac{a_{\rm res}}{\left[2(B-B_{\rm
res})/\Gamma_B^{\rm inel}\right]^2+1}. \label{eqsiginel}
\end{eqnarray}
The resulting profiles are shown in Figs.\ \ref{figxs2} and
\ref{figsl2}, using the parameters determined from the $S$ matrix
elements without refitting to the scattering lengths and cross
sections.

It is remarkable that, even in a system as weakly coupled as
He-NH, the behavior of the cross sections and scattering lengths
is so different from the elastic case. It has been common in
theoretical studies of atom-atom scattering to model Feshbach
resonances using a 2-channel treatment with one closed and one
open channel. The present results make it clear that such an
approximation may miss essential features of the physics, and in
particular may predict unphysical poles in the scattering length.

Feshbach resonances with $L>0$ in the incoming channel can also
occur, but are suppressed at ultralow energies because of the
centrifugal barrier. $^3$He-NH has a very low reduced mass, and
with a long-range potential given by $-C_6 R^{-6}$ with $C_6 =
7.84 E_h\ a_0^6$ the heights and positions of the centrifugal
barriers for $L=1$ to $4$ are given in Table \ref{tab:centdata}.
Because of this, no zero-energy resonance was observed for
$M_{\cal J}=-1$ near 14345 G, even though there is a state that
crosses the threshold there.

\begin{table}[tb]
\caption{Positions (\AA) and heights (K) of centrifugal barriers.
\label{tab:centdata}}
\begin{tabular}{ccccccccc}\hline\hline
      &&       \multicolumn{3}{c}{\rm $^3$He-NH}     &&      \multicolumn{3}{c}{\rm $^4$He-NH}      \\ \cline{3-5}\cline{7-9}
  $L$ &&   $R_{\rm max}$ && $V(R_{\rm max})$ &&  $R_{\rm max}$ && $V(R_{\rm max})$ \\ \hline
   1  &&    9.6    &&    0.14    &&    10.2   &&   0.10   \\
   2  &&    7.3    &&    0.73    &&    7.7    &&   0.52   \\
   3  &&    6.1    &&    2.06    &&    6.5    &&   1.46   \\
   4  &&    5.4    &&    4.43    &&    5.7    &&   3.14   \\ \hline\hline
\end{tabular}
\end{table}

We should emphasize the advantage of using coupled-channel methods
rather than basis-set methods for bound states when attempting to
locate zero-energy Feshbach resonances. Basis-set methods lose
accuracy close to dissociation because of the difficulty of
representing near-dissociation and continuum functions with basis
sets. In coupled-channel methods, by contrast, the behavior of the
wavefunction at long range can be built in by applying WKB
boundary conditions at $R_{\rm max}$. Because of this, the
approximations in the bound-state calculations are very similar to
those in scattering calculations with the same basis set, and as
we have shown this allows very precise estimates of resonance
positions.

\section{Relative merits of different basis sets}

The coupled and uncoupled basis sets give identical results for
bound-state energies and for collision properties, and in this
sense they are equivalent. However, the uncoupled basis set is a
little easier to program and its matrix elements are easier to
generalize to more complicated cases (such as those involving
nuclear spin or two structured monomers). In addition, the
uncoupled basis set gives a much simpler representation of the
wavefunctions for bound states and resonances in any significant
magnetic field. We therefore intend to use uncoupled basis sets in
future work.

\section{Conclusions}

We have modified the BOUND and MOLSCAT packages to allow the use
of basis sets in which the asymptotic Hamiltonian is non-diagonal
and used the new capability to perform bound-state and scattering
calculations on He-NH in the presence of a magnetic field. The
bound-state capability makes it possible to locate zero-energy
Feshbach resonances as a function of magnetic field even when they
are very narrow. The new capability provides a very
straightforward way to program new coupling cases and collision
types, and in future work we will use it to investigate low-energy
scattering of systems containing two structured monomers in
electric and magnetic fields.

For He-NH, we have located two zero-energy Feshbach resonances
involving an $m_s=+1$ level of the He-NH complex tuned through the
$m_s=-1$ and $m_s=0$ thresholds. The two resonances show very
different behavior as a function of magnetic field. For the
resonance at the $m_s=-1$ threshold, only elastic scattering is
possible and the resonance shows the classic behavior familiar
from ultracold atom-atom scattering. The scattering length passes
through a pole at resonance and the elastic cross section shows a
very large peak. For the resonance at the $m_s=0$ threshold,
however, inelastic scattering is also possible. The pole in the
scattering length is dramatically suppressed, and the peak in the
elastic cross section is far smaller than would be expected if a
pole were present.

Our results provide a numerical demonstration of the effects
recently predicted by Hutson \cite{Hutson:res:2006}, who
parameterized the strength of the resonant contribution with a
resonant scattering length $a_{\rm res}$. For the inelastic
resonance that we have characterized in He-NH, the resonant
scattering length is only 8.96~\AA.

The suppression of resonant peaks in cross sections by inelastic
scattering will be a very general effect. Most molecular systems
have stronger inelasticity than He-NH, and will have even smaller
resonant scattering lengths. In such cases the peaks will be even
more strongly suppressed than here. The effect explains previously
puzzling results in Na + Na$_2$ \cite{Quemener:2004} and Li +
Li$_2$ \cite{Cvitas:li3:2006} scattering, where it was found that
cross sections for ultracold collisions of vibrationally excited
molecules were only weakly dependent on potential parameters and
showed no sharp peaks when Feshbach resonances were tuned through
zero energy. The lack of such peaks is now seen to be due to
suppression by inelastic processes. This suggests that
calculations on ultracold collisions of molecules in the presence
of inelasticity may be much less sensitive to potential details
than was previously expected.

\acknowledgments

The authors are grateful to the Royal Society for an International
Joint Project grant which made this collaboration possible and to
Marko Cvita\v{s} and Pavel Sold\'{a}n for comments on the
manuscript.

\bibliography{../../all}

\begin{thebibliography}{68}
\expandafter\ifx\csname natexlab\endcsname\relax\def\natexlab#1{#1}\fi
\expandafter\ifx\csname bibnamefont\endcsname\relax
  \def\bibnamefont#1{#1}\fi
\expandafter\ifx\csname bibfnamefont\endcsname\relax
  \def\bibfnamefont#1{#1}\fi
\expandafter\ifx\csname citenamefont\endcsname\relax
  \def\citenamefont#1{#1}\fi
\expandafter\ifx\csname url\endcsname\relax
  \def\url#1{\texttt{#1}}\fi
\expandafter\ifx\csname urlprefix\endcsname\relax\def\urlprefix{URL }\fi
\providecommand{\bibinfo}[2]{#2}
\providecommand{\eprint}[2][]{\url{#2}}

\bibitem[{\citenamefont{Hutson and Sold\'{a}n}(2006)}]{Hutson:IRPC:2006}
\bibinfo{author}{\bibfnamefont{J.~M.} \bibnamefont{Hutson}} \bibnamefont{and}
  \bibinfo{author}{\bibfnamefont{P.}~\bibnamefont{Sold\'{a}n}},
  \bibinfo{journal}{Int. Rev. Phys. Chem.} \textbf{\bibinfo{volume}{25}},
  \bibinfo{pages}{497} (\bibinfo{year}{2006}).

\bibitem[{\citenamefont{K\"{o}hler et~al.}(2006)\citenamefont{K\"{o}hler,
  Goral, and Julienne}}]{Koehler:RMP:2006}
\bibinfo{author}{\bibfnamefont{T.}~\bibnamefont{K\"{o}hler}},
  \bibinfo{author}{\bibfnamefont{K.}~\bibnamefont{Goral}}, \bibnamefont{and}
  \bibinfo{author}{\bibfnamefont{P.~S.} \bibnamefont{Julienne}},
  \bibinfo{journal}{Rev. Mod. Phys.} \textbf{\bibinfo{volume}{78}},
  \bibinfo{pages}{000} (\bibinfo{year}{2006}).

\bibitem[{\citenamefont{Roberts et~al.}(2001)\citenamefont{Roberts, Claussen,
  Cornish, Donley, Cornell, and Wieman}}]{Roberts:2001}
\bibinfo{author}{\bibfnamefont{J.~L.} \bibnamefont{Roberts}},
  \bibinfo{author}{\bibfnamefont{N.~R.} \bibnamefont{Claussen}},
  \bibinfo{author}{\bibfnamefont{S.~L.} \bibnamefont{Cornish}},
  \bibinfo{author}{\bibfnamefont{E.~A.} \bibnamefont{Donley}},
  \bibinfo{author}{\bibfnamefont{E.~A.} \bibnamefont{Cornell}},
  \bibnamefont{and} \bibinfo{author}{\bibfnamefont{C.~E.}
  \bibnamefont{Wieman}}, \bibinfo{journal}{Phys. Rev. Lett.}
  \textbf{\bibinfo{volume}{86}}, \bibinfo{pages}{4211} (\bibinfo{year}{2001}).

\bibitem[{\citenamefont{Donley et~al.}(2002)\citenamefont{Donley, Claussen,
  Thompson, and Wieman}}]{Donley:2002}
\bibinfo{author}{\bibfnamefont{E.~A.} \bibnamefont{Donley}},
  \bibinfo{author}{\bibfnamefont{N.~R.} \bibnamefont{Claussen}},
  \bibinfo{author}{\bibfnamefont{S.~T.} \bibnamefont{Thompson}},
  \bibnamefont{and} \bibinfo{author}{\bibfnamefont{C.~E.}
  \bibnamefont{Wieman}}, \bibinfo{journal}{Nature}
  \textbf{\bibinfo{volume}{417}}, \bibinfo{pages}{529} (\bibinfo{year}{2002}).

\bibitem[{\citenamefont{Herbig et~al.}(2003)\citenamefont{Herbig, Kraemer,
  Mark, Weber, Chin, N\"{a}gerl, and Grimm}}]{Herbig:2003}
\bibinfo{author}{\bibfnamefont{J.}~\bibnamefont{Herbig}},
  \bibinfo{author}{\bibfnamefont{T.}~\bibnamefont{Kraemer}},
  \bibinfo{author}{\bibfnamefont{M.}~\bibnamefont{Mark}},
  \bibinfo{author}{\bibfnamefont{T.}~\bibnamefont{Weber}},
  \bibinfo{author}{\bibfnamefont{C.}~\bibnamefont{Chin}},
  \bibinfo{author}{\bibfnamefont{H.~C.} \bibnamefont{N\"{a}gerl}},
  \bibnamefont{and} \bibinfo{author}{\bibfnamefont{R.}~\bibnamefont{Grimm}},
  \bibinfo{journal}{Science} \textbf{\bibinfo{volume}{301}},
  \bibinfo{pages}{1510} (\bibinfo{year}{2003}).

\bibitem[{\citenamefont{Xu et~al.}(2003)\citenamefont{Xu, Mukaiyama,
  Abo-Shaeer, Chin, Miller, and Ketterle}}]{Xu:2003}
\bibinfo{author}{\bibfnamefont{K.}~\bibnamefont{Xu}},
  \bibinfo{author}{\bibfnamefont{T.}~\bibnamefont{Mukaiyama}},
  \bibinfo{author}{\bibfnamefont{J.~R.} \bibnamefont{Abo-Shaeer}},
  \bibinfo{author}{\bibfnamefont{J.~K.} \bibnamefont{Chin}},
  \bibinfo{author}{\bibfnamefont{D.~E.} \bibnamefont{Miller}},
  \bibnamefont{and} \bibinfo{author}{\bibfnamefont{W.}~\bibnamefont{Ketterle}},
  \bibinfo{journal}{Phys. Rev. Lett.} \textbf{\bibinfo{volume}{91}},
  \bibinfo{pages}{210402} (\bibinfo{year}{2003}).

\bibitem[{\citenamefont{D\"urr et~al.}(2004)\citenamefont{D\"urr, Volz, Marte,
  and Rempe}}]{Durr:mol87Rb:2004}
\bibinfo{author}{\bibfnamefont{S.}~\bibnamefont{D\"urr}},
  \bibinfo{author}{\bibfnamefont{T.}~\bibnamefont{Volz}},
  \bibinfo{author}{\bibfnamefont{A.}~\bibnamefont{Marte}}, \bibnamefont{and}
  \bibinfo{author}{\bibfnamefont{G.}~\bibnamefont{Rempe}},
  \bibinfo{journal}{Phys. Rev. Lett.} \textbf{\bibinfo{volume}{92}},
  \bibinfo{pages}{020406} (\bibinfo{year}{2004}).

\bibitem[{\citenamefont{Regal et~al.}(2003)\citenamefont{Regal, Ticknor, Bohn,
  and Jin}}]{Regal:40K2:2003}
\bibinfo{author}{\bibfnamefont{C.~A.} \bibnamefont{Regal}},
  \bibinfo{author}{\bibfnamefont{C.}~\bibnamefont{Ticknor}},
  \bibinfo{author}{\bibfnamefont{J.~L.} \bibnamefont{Bohn}}, \bibnamefont{and}
  \bibinfo{author}{\bibfnamefont{D.~S.} \bibnamefont{Jin}},
  \bibinfo{journal}{Nature} \textbf{\bibinfo{volume}{424}}, \bibinfo{pages}{47}
  (\bibinfo{year}{2003}).

\bibitem[{\citenamefont{Strecker et~al.}(2003)\citenamefont{Strecker,
  Partridge, and Hulet}}]{Strecker:2003}
\bibinfo{author}{\bibfnamefont{K.~E.} \bibnamefont{Strecker}},
  \bibinfo{author}{\bibfnamefont{G.~B.} \bibnamefont{Partridge}},
  \bibnamefont{and} \bibinfo{author}{\bibfnamefont{R.~G.} \bibnamefont{Hulet}},
  \bibinfo{journal}{Phys. Rev. Lett.} \textbf{\bibinfo{volume}{91}},
  \bibinfo{pages}{080406} (\bibinfo{year}{2003}).

\bibitem[{\citenamefont{Cubizolles et~al.}(2003)\citenamefont{Cubizolles,
  Bourdel, Kokkelmans, Shlyapnikov, and Salomon}}]{Cubizolles:2003}
\bibinfo{author}{\bibfnamefont{J.}~\bibnamefont{Cubizolles}},
  \bibinfo{author}{\bibfnamefont{T.}~\bibnamefont{Bourdel}},
  \bibinfo{author}{\bibfnamefont{S.~J. J. M.~F.} \bibnamefont{Kokkelmans}},
  \bibinfo{author}{\bibfnamefont{G.~V.} \bibnamefont{Shlyapnikov}},
  \bibnamefont{and} \bibinfo{author}{\bibfnamefont{C.}~\bibnamefont{Salomon}},
  \bibinfo{journal}{Phys. Rev. Lett.} \textbf{\bibinfo{volume}{91}},
  \bibinfo{pages}{240401} (\bibinfo{year}{2003}).

\bibitem[{\citenamefont{Jochim et~al.}(2003{\natexlab{a}})\citenamefont{Jochim,
  Bartenstein, Altmeyer, Hendl, Chin, Denschlag, and
  Grimm}}]{Jochim:Li2pure:2003}
\bibinfo{author}{\bibfnamefont{S.}~\bibnamefont{Jochim}},
  \bibinfo{author}{\bibfnamefont{M.}~\bibnamefont{Bartenstein}},
  \bibinfo{author}{\bibfnamefont{A.}~\bibnamefont{Altmeyer}},
  \bibinfo{author}{\bibfnamefont{G.}~\bibnamefont{Hendl}},
  \bibinfo{author}{\bibfnamefont{C.}~\bibnamefont{Chin}},
  \bibinfo{author}{\bibfnamefont{J.~H.} \bibnamefont{Denschlag}},
  \bibnamefont{and} \bibinfo{author}{\bibfnamefont{R.}~\bibnamefont{Grimm}},
  \bibinfo{journal}{Phys. Rev. Lett.} \textbf{\bibinfo{volume}{91}},
  \bibinfo{pages}{240402} (\bibinfo{year}{2003}{\natexlab{a}}).

\bibitem[{\citenamefont{Jochim et~al.}(2003{\natexlab{b}})\citenamefont{Jochim,
  Bartenstein, Altmeyer, Hendl, Riedl, Chin, Denschlag, and
  Grimm}}]{Jochim:Li2BEC:2003}
\bibinfo{author}{\bibfnamefont{S.}~\bibnamefont{Jochim}},
  \bibinfo{author}{\bibfnamefont{M.}~\bibnamefont{Bartenstein}},
  \bibinfo{author}{\bibfnamefont{A.}~\bibnamefont{Altmeyer}},
  \bibinfo{author}{\bibfnamefont{G.}~\bibnamefont{Hendl}},
  \bibinfo{author}{\bibfnamefont{S.}~\bibnamefont{Riedl}},
  \bibinfo{author}{\bibfnamefont{C.}~\bibnamefont{Chin}},
  \bibinfo{author}{\bibfnamefont{J.~H.} \bibnamefont{Denschlag}},
  \bibnamefont{and} \bibinfo{author}{\bibfnamefont{R.}~\bibnamefont{Grimm}},
  \bibinfo{journal}{Science} \textbf{\bibinfo{volume}{302}},
  \bibinfo{pages}{2101} (\bibinfo{year}{2003}{\natexlab{b}}).

\bibitem[{\citenamefont{Zwierlein et~al.}(2003)\citenamefont{Zwierlein, Stan,
  Schunck, Raupach, Gupta, Hadzibabic, and Ketterle}}]{Zwierlein:2003}
\bibinfo{author}{\bibfnamefont{M.~W.} \bibnamefont{Zwierlein}},
  \bibinfo{author}{\bibfnamefont{C.~A.} \bibnamefont{Stan}},
  \bibinfo{author}{\bibfnamefont{C.~H.} \bibnamefont{Schunck}},
  \bibinfo{author}{\bibfnamefont{S.~M.~F.} \bibnamefont{Raupach}},
  \bibinfo{author}{\bibfnamefont{S.}~\bibnamefont{Gupta}},
  \bibinfo{author}{\bibfnamefont{Z.}~\bibnamefont{Hadzibabic}},
  \bibnamefont{and} \bibinfo{author}{\bibfnamefont{W.}~\bibnamefont{Ketterle}},
  \bibinfo{journal}{Phys. Rev. Lett.} \textbf{\bibinfo{volume}{91}},
  \bibinfo{pages}{250401} (\bibinfo{year}{2003}).

\bibitem[{\citenamefont{Greiner et~al.}(2003)\citenamefont{Greiner, Regal, and
  Jin}}]{Greiner:2003}
\bibinfo{author}{\bibfnamefont{M.}~\bibnamefont{Greiner}},
  \bibinfo{author}{\bibfnamefont{C.~A.} \bibnamefont{Regal}}, \bibnamefont{and}
  \bibinfo{author}{\bibfnamefont{D.~S.} \bibnamefont{Jin}},
  \bibinfo{journal}{Nature} \textbf{\bibinfo{volume}{426}},
  \bibinfo{pages}{537} (\bibinfo{year}{2003}).

\bibitem[{\citenamefont{Kraemer et~al.}(2006)\citenamefont{Kraemer, Mark,
  Waldburger, Danzl, Chin, Engeser, Lange, Pilch, Jaakkola, N\"{a}gerl
  et~al.}}]{Kraemer:2006}
\bibinfo{author}{\bibfnamefont{T.}~\bibnamefont{Kraemer}},
  \bibinfo{author}{\bibfnamefont{M.}~\bibnamefont{Mark}},
  \bibinfo{author}{\bibfnamefont{P.}~\bibnamefont{Waldburger}},
  \bibinfo{author}{\bibfnamefont{J.~G.} \bibnamefont{Danzl}},
  \bibinfo{author}{\bibfnamefont{C.}~\bibnamefont{Chin}},
  \bibinfo{author}{\bibfnamefont{B.}~\bibnamefont{Engeser}},
  \bibinfo{author}{\bibfnamefont{A.~D.} \bibnamefont{Lange}},
  \bibinfo{author}{\bibfnamefont{K.}~\bibnamefont{Pilch}},
  \bibinfo{author}{\bibfnamefont{A.}~\bibnamefont{Jaakkola}},
  \bibinfo{author}{\bibfnamefont{H.~C.} \bibnamefont{N\"{a}gerl}},
  \bibnamefont{et~al.}, \bibinfo{journal}{Nature}
  \textbf{\bibinfo{volume}{440}}, \bibinfo{pages}{315} (\bibinfo{year}{2006}).

\bibitem[{\citenamefont{Chin et~al.}(2005)\citenamefont{Chin, Kraemer, Mark,
  Herbig, Waldburger, N\"{a}gerl, and Grimm}}]{Chin:2005}
\bibinfo{author}{\bibfnamefont{C.}~\bibnamefont{Chin}},
  \bibinfo{author}{\bibfnamefont{T.}~\bibnamefont{Kraemer}},
  \bibinfo{author}{\bibfnamefont{M.}~\bibnamefont{Mark}},
  \bibinfo{author}{\bibfnamefont{J.}~\bibnamefont{Herbig}},
  \bibinfo{author}{\bibfnamefont{P.}~\bibnamefont{Waldburger}},
  \bibinfo{author}{\bibfnamefont{H.~C.} \bibnamefont{N\"{a}gerl}},
  \bibnamefont{and} \bibinfo{author}{\bibfnamefont{R.}~\bibnamefont{Grimm}},
  \bibinfo{journal}{Phys. Rev. Lett.} \textbf{\bibinfo{volume}{94}},
  \bibinfo{pages}{123201} (\bibinfo{year}{2005}).

\bibitem[{\citenamefont{Bartenstein et~al.}(2004)\citenamefont{Bartenstein,
  Altmeyer, Riedl, Jochim, Chin, Denschlag, and
  Grimm}}]{Bartenstein:crossover:2004}
\bibinfo{author}{\bibfnamefont{M.}~\bibnamefont{Bartenstein}},
  \bibinfo{author}{\bibfnamefont{A.}~\bibnamefont{Altmeyer}},
  \bibinfo{author}{\bibfnamefont{S.}~\bibnamefont{Riedl}},
  \bibinfo{author}{\bibfnamefont{S.}~\bibnamefont{Jochim}},
  \bibinfo{author}{\bibfnamefont{C.}~\bibnamefont{Chin}},
  \bibinfo{author}{\bibfnamefont{J.~H.} \bibnamefont{Denschlag}},
  \bibnamefont{and} \bibinfo{author}{\bibfnamefont{R.}~\bibnamefont{Grimm}},
  \bibinfo{journal}{Phys. Rev. Lett.} \textbf{\bibinfo{volume}{92}},
  \bibinfo{pages}{120401} (\bibinfo{year}{2004}).

\bibitem[{\citenamefont{Regal et~al.}(2004)\citenamefont{Regal, Greiner, and
  Jin}}]{Regal:res-cond:2004}
\bibinfo{author}{\bibfnamefont{C.~A.} \bibnamefont{Regal}},
  \bibinfo{author}{\bibfnamefont{M.}~\bibnamefont{Greiner}}, \bibnamefont{and}
  \bibinfo{author}{\bibfnamefont{D.~S.} \bibnamefont{Jin}},
  \bibinfo{journal}{Phys. Rev. Lett.} \textbf{\bibinfo{volume}{92}},
  \bibinfo{pages}{040403} (\bibinfo{year}{2004}).

\bibitem[{\citenamefont{Zwierlein et~al.}(2004)\citenamefont{Zwierlein, Stan,
  Schunck, Raupach, Kerman, and Ketterle}}]{Zwierlein:2004}
\bibinfo{author}{\bibfnamefont{M.~W.} \bibnamefont{Zwierlein}},
  \bibinfo{author}{\bibfnamefont{C.~A.} \bibnamefont{Stan}},
  \bibinfo{author}{\bibfnamefont{C.~H.} \bibnamefont{Schunck}},
  \bibinfo{author}{\bibfnamefont{S.~M.~F.} \bibnamefont{Raupach}},
  \bibinfo{author}{\bibfnamefont{A.~J.} \bibnamefont{Kerman}},
  \bibnamefont{and} \bibinfo{author}{\bibfnamefont{W.}~\bibnamefont{Ketterle}},
  \bibinfo{journal}{Phys. Rev. Lett.} \textbf{\bibinfo{volume}{92}},
  \bibinfo{pages}{120403} (\bibinfo{year}{2004}).

\bibitem[{\citenamefont{Weinstein et~al.}(1998)\citenamefont{Weinstein,
  deCarvalho, Guillet, Friedrich, and Doyle}}]{Weinstein:CaH:1998}
\bibinfo{author}{\bibfnamefont{J.~D.} \bibnamefont{Weinstein}},
  \bibinfo{author}{\bibfnamefont{R.}~\bibnamefont{deCarvalho}},
  \bibinfo{author}{\bibfnamefont{T.}~\bibnamefont{Guillet}},
  \bibinfo{author}{\bibfnamefont{B.}~\bibnamefont{Friedrich}},
  \bibnamefont{and} \bibinfo{author}{\bibfnamefont{J.~M.} \bibnamefont{Doyle}},
  \bibinfo{journal}{Nature} \textbf{\bibinfo{volume}{395}},
  \bibinfo{pages}{148} (\bibinfo{year}{1998}).

\bibitem[{\citenamefont{Egorov et~al.}(2004)\citenamefont{Egorov, Campbell,
  Friedrich, Maxwell, Tsikata, van Buuren, and Doyle}}]{Egorov:2004}
\bibinfo{author}{\bibfnamefont{D.}~\bibnamefont{Egorov}},
  \bibinfo{author}{\bibfnamefont{W.~C.} \bibnamefont{Campbell}},
  \bibinfo{author}{\bibfnamefont{B.}~\bibnamefont{Friedrich}},
  \bibinfo{author}{\bibfnamefont{S.~E.} \bibnamefont{Maxwell}},
  \bibinfo{author}{\bibfnamefont{E.}~\bibnamefont{Tsikata}},
  \bibinfo{author}{\bibfnamefont{L.~D.} \bibnamefont{van Buuren}},
  \bibnamefont{and} \bibinfo{author}{\bibfnamefont{J.~M.} \bibnamefont{Doyle}},
  \bibinfo{journal}{Eur. Phys. J. D} \textbf{\bibinfo{volume}{31}},
  \bibinfo{pages}{307} (\bibinfo{year}{2004}).

\bibitem[{\citenamefont{Bethlem and Meijer}(2003)}]{Bethlem:IRPC:2003}
\bibinfo{author}{\bibfnamefont{H.~L.} \bibnamefont{Bethlem}} \bibnamefont{and}
  \bibinfo{author}{\bibfnamefont{G.}~\bibnamefont{Meijer}},
  \bibinfo{journal}{Int. Rev. Phys. Chem.} \textbf{\bibinfo{volume}{22}},
  \bibinfo{pages}{73} (\bibinfo{year}{2003}).

\bibitem[{\citenamefont{Bethlem et~al.}(2006)\citenamefont{Bethlem, Tarbutt,
  K\"{u}pper, Carty, Wohlfart, Hinds, and Meijer}}]{Bethlem:2006}
\bibinfo{author}{\bibfnamefont{H.~L.} \bibnamefont{Bethlem}},
  \bibinfo{author}{\bibfnamefont{M.~R.} \bibnamefont{Tarbutt}},
  \bibinfo{author}{\bibfnamefont{J.}~\bibnamefont{K\"{u}pper}},
  \bibinfo{author}{\bibfnamefont{D.}~\bibnamefont{Carty}},
  \bibinfo{author}{\bibfnamefont{K.}~\bibnamefont{Wohlfart}},
  \bibinfo{author}{\bibfnamefont{E.~A.} \bibnamefont{Hinds}}, \bibnamefont{and}
  \bibinfo{author}{\bibfnamefont{G.}~\bibnamefont{Meijer}},
  \bibinfo{journal}{J. Phys. B-At. Mol. Opt. Phys.}
  \textbf{\bibinfo{volume}{39}}, \bibinfo{pages}{R263} (\bibinfo{year}{2006}).

\bibitem[{\citenamefont{Domokos and Ritsch}(2002)}]{Domokos:2002}
\bibinfo{author}{\bibfnamefont{P.}~\bibnamefont{Domokos}} \bibnamefont{and}
  \bibinfo{author}{\bibfnamefont{H.}~\bibnamefont{Ritsch}},
  \bibinfo{journal}{Phys. Rev. Lett.} \textbf{\bibinfo{volume}{89}},
  \bibinfo{pages}{253003} (\bibinfo{year}{2002}).

\bibitem[{\citenamefont{Chan et~al.}(2003)\citenamefont{Chan, Black, and
  Vuletic}}]{Chan:2003}
\bibinfo{author}{\bibfnamefont{H.~W.} \bibnamefont{Chan}},
  \bibinfo{author}{\bibfnamefont{A.~T.} \bibnamefont{Black}}, \bibnamefont{and}
  \bibinfo{author}{\bibfnamefont{V.}~\bibnamefont{Vuletic}},
  \bibinfo{journal}{Phys. Rev. Lett.} \textbf{\bibinfo{volume}{90}},
  \bibinfo{pages}{063003} (\bibinfo{year}{2003}).

\bibitem[{\citenamefont{Krems}(2005)}]{Krems:IRPC:2005}
\bibinfo{author}{\bibfnamefont{R.~V.} \bibnamefont{Krems}},
  \bibinfo{journal}{Int. Rev. Phys. Chem.} \textbf{\bibinfo{volume}{24}},
  \bibinfo{pages}{99} (\bibinfo{year}{2005}).

\bibitem[{\citenamefont{Volpi and Bohn}(2002)}]{Volpi:2002}
\bibinfo{author}{\bibfnamefont{A.}~\bibnamefont{Volpi}} \bibnamefont{and}
  \bibinfo{author}{\bibfnamefont{J.~L.} \bibnamefont{Bohn}},
  \bibinfo{journal}{Phys. Rev. A} \textbf{\bibinfo{volume}{65}},
  \bibinfo{pages}{052712} (\bibinfo{year}{2002}).

\bibitem[{\citenamefont{Krems et~al.}(2003)\citenamefont{Krems, Sadeghpour,
  Dalgarno, Zgid, K{\l}os, and Cha{\l}asi\'{n}ski}}]{Krems:henh:2003}
\bibinfo{author}{\bibfnamefont{R.~V.} \bibnamefont{Krems}},
  \bibinfo{author}{\bibfnamefont{H.~R.} \bibnamefont{Sadeghpour}},
  \bibinfo{author}{\bibfnamefont{A.}~\bibnamefont{Dalgarno}},
  \bibinfo{author}{\bibfnamefont{D.}~\bibnamefont{Zgid}},
  \bibinfo{author}{\bibfnamefont{J.}~\bibnamefont{K{\l}os}}, \bibnamefont{and}
  \bibinfo{author}{\bibfnamefont{G.}~\bibnamefont{Cha{\l}asi\'{n}ski}},
  \bibinfo{journal}{Phys. Rev. A} \textbf{\bibinfo{volume}{68}},
  \bibinfo{pages}{051401} (\bibinfo{year}{2003}).

\bibitem[{\citenamefont{Cybulski et~al.}(2005)\citenamefont{Cybulski, Krems,
  Sadeghpour, Dalgarno, K{\l}os, Groenenboom, van~der Avoird, Zgid, and
  Cha{\l}asi\'{n}ski}}]{Cybulski:2005}
\bibinfo{author}{\bibfnamefont{H.}~\bibnamefont{Cybulski}},
  \bibinfo{author}{\bibfnamefont{R.~V.} \bibnamefont{Krems}},
  \bibinfo{author}{\bibfnamefont{H.~R.} \bibnamefont{Sadeghpour}},
  \bibinfo{author}{\bibfnamefont{A.}~\bibnamefont{Dalgarno}},
  \bibinfo{author}{\bibfnamefont{J.}~\bibnamefont{K{\l}os}},
  \bibinfo{author}{\bibfnamefont{G.~C.} \bibnamefont{Groenenboom}},
  \bibinfo{author}{\bibfnamefont{A.}~\bibnamefont{van~der Avoird}},
  \bibinfo{author}{\bibfnamefont{D.}~\bibnamefont{Zgid}}, \bibnamefont{and}
  \bibinfo{author}{\bibfnamefont{G.}~\bibnamefont{Cha{\l}asi\'{n}ski}},
  \bibinfo{journal}{J. Chem. Phys.} \textbf{\bibinfo{volume}{122}},
  \bibinfo{pages}{094307} (\bibinfo{year}{2005}).

\bibitem[{\citenamefont{Krems and
  Dalgarno}(2004{\natexlab{a}})}]{Krems:mfield:2004}
\bibinfo{author}{\bibfnamefont{R.~V.} \bibnamefont{Krems}} \bibnamefont{and}
  \bibinfo{author}{\bibfnamefont{A.}~\bibnamefont{Dalgarno}},
  \bibinfo{journal}{J. Chem. Phys.} \textbf{\bibinfo{volume}{120}},
  \bibinfo{pages}{2296} (\bibinfo{year}{2004}{\natexlab{a}}).

\bibitem[{\citenamefont{Krems and
  Dalgarno}(2004{\natexlab{b}})}]{Krems:FWQC:2004}
\bibinfo{author}{\bibfnamefont{R.~V.} \bibnamefont{Krems}} \bibnamefont{and}
  \bibinfo{author}{\bibfnamefont{A.}~\bibnamefont{Dalgarno}}, in
  \emph{\bibinfo{booktitle}{Fundamental World of Quantum Chemistry}}, edited by
  \bibinfo{editor}{\bibfnamefont{E.~J.} \bibnamefont{Br{\"a}ndas}}
  \bibnamefont{and} \bibinfo{editor}{\bibfnamefont{E.~S.}
  \bibnamefont{Kryachko}} (\bibinfo{publisher}{Kluwer Academic},
  \bibinfo{year}{2004}{\natexlab{b}}), vol.~\bibinfo{volume}{3}, pp.
  \bibinfo{pages}{273--294}.

\bibitem[{\citenamefont{Ticknor and
  Bohn}(2005{\natexlab{a}})}]{Ticknor:OHmag:2005}
\bibinfo{author}{\bibfnamefont{C.}~\bibnamefont{Ticknor}} \bibnamefont{and}
  \bibinfo{author}{\bibfnamefont{J.~L.} \bibnamefont{Bohn}},
  \bibinfo{journal}{Phys. Rev. A} \textbf{\bibinfo{volume}{71}},
  \bibinfo{pages}{022709} (\bibinfo{year}{2005}{\natexlab{a}}).

\bibitem[{\citenamefont{Lara et~al.}(2006{\natexlab{a}})\citenamefont{Lara,
  Bohn, Potter, Sold\'{a}n, and Hutson}}]{Lara:PRL:2006}
\bibinfo{author}{\bibfnamefont{M.}~\bibnamefont{Lara}},
  \bibinfo{author}{\bibfnamefont{J.~L.} \bibnamefont{Bohn}},
  \bibinfo{author}{\bibfnamefont{D.~E.} \bibnamefont{Potter}},
  \bibinfo{author}{\bibfnamefont{P.}~\bibnamefont{Sold\'{a}n}},
  \bibnamefont{and} \bibinfo{author}{\bibfnamefont{J.~M.}
  \bibnamefont{Hutson}}, \bibinfo{journal}{Phys. Rev. Lett., in press, and
  arXiv:physics/0607084}  (\bibinfo{year}{2006}{\natexlab{a}}).

\bibitem[{\citenamefont{Lara et~al.}(2006{\natexlab{b}})\citenamefont{Lara,
  Bohn, Potter, Sold\'{a}n, and Hutson}}]{Lara:PRA:2006}
\bibinfo{author}{\bibfnamefont{M.}~\bibnamefont{Lara}},
  \bibinfo{author}{\bibfnamefont{J.~L.} \bibnamefont{Bohn}},
  \bibinfo{author}{\bibfnamefont{D.~E.} \bibnamefont{Potter}},
  \bibinfo{author}{\bibfnamefont{P.}~\bibnamefont{Sold\'{a}n}},
  \bibnamefont{and} \bibinfo{author}{\bibfnamefont{J.~M.}
  \bibnamefont{Hutson}}, \bibinfo{journal}{Phys. Rev. A, in press, and
  arXiv:physics/0608200}  (\bibinfo{year}{2006}{\natexlab{b}}).

\bibitem[{\citenamefont{Avdeenkov and Bohn}(2002)}]{Avdeenkov:2002}
\bibinfo{author}{\bibfnamefont{A.~V.} \bibnamefont{Avdeenkov}}
  \bibnamefont{and} \bibinfo{author}{\bibfnamefont{J.~L.} \bibnamefont{Bohn}},
  \bibinfo{journal}{Phys. Rev. A} \textbf{\bibinfo{volume}{66}},
  \bibinfo{pages}{052718} (\bibinfo{year}{2002}).

\bibitem[{\citenamefont{Avdeenkov and Bohn}(2003)}]{Avdeenkov:2003}
\bibinfo{author}{\bibfnamefont{A.~V.} \bibnamefont{Avdeenkov}}
  \bibnamefont{and} \bibinfo{author}{\bibfnamefont{J.~L.} \bibnamefont{Bohn}},
  \bibinfo{journal}{Phys. Rev. Lett.} \textbf{\bibinfo{volume}{90}},
  \bibinfo{pages}{043006} (\bibinfo{year}{2003}).

\bibitem[{\citenamefont{Avdeenkov et~al.}(2004)\citenamefont{Avdeenkov,
  Bortolotti, and Bohn}}]{Avdeenkov:2004}
\bibinfo{author}{\bibfnamefont{A.~V.} \bibnamefont{Avdeenkov}},
  \bibinfo{author}{\bibfnamefont{D.~C.~E.} \bibnamefont{Bortolotti}},
  \bibnamefont{and} \bibinfo{author}{\bibfnamefont{J.~L.} \bibnamefont{Bohn}},
  \bibinfo{journal}{Phys. Rev. A} \textbf{\bibinfo{volume}{69}},
  \bibinfo{pages}{012710} (\bibinfo{year}{2004}).

\bibitem[{\citenamefont{Avdeenkov and Bohn}(2005)}]{Avdeenkov:2005}
\bibinfo{author}{\bibfnamefont{A.~V.} \bibnamefont{Avdeenkov}}
  \bibnamefont{and} \bibinfo{author}{\bibfnamefont{J.~L.} \bibnamefont{Bohn}},
  \bibinfo{journal}{Phys. Rev. A} \textbf{\bibinfo{volume}{71}},
  \bibinfo{pages}{022706} (\bibinfo{year}{2005}).

\bibitem[{\citenamefont{Ticknor and
  Bohn}(2005{\natexlab{b}})}]{Ticknor:long-range:2005}
\bibinfo{author}{\bibfnamefont{C.}~\bibnamefont{Ticknor}} \bibnamefont{and}
  \bibinfo{author}{\bibfnamefont{J.~L.} \bibnamefont{Bohn}},
  \bibinfo{journal}{Phys. Rev. A} \textbf{\bibinfo{volume}{72}},
  \bibinfo{pages}{032717} (\bibinfo{year}{2005}{\natexlab{b}}).

\bibitem[{\citenamefont{Avdeenkov et~al.}(2006)\citenamefont{Avdeenkov, Kajita,
  and Bohn}}]{Avdeenkov:2006}
\bibinfo{author}{\bibfnamefont{A.~V.} \bibnamefont{Avdeenkov}},
  \bibinfo{author}{\bibfnamefont{M.}~\bibnamefont{Kajita}}, \bibnamefont{and}
  \bibinfo{author}{\bibfnamefont{J.~L.} \bibnamefont{Bohn}},
  \bibinfo{journal}{Phys. Rev. A} \textbf{\bibinfo{volume}{73}},
  \bibinfo{pages}{022707} (\bibinfo{year}{2006}).

\bibitem[{\citenamefont{Tscherbul and Krems}(2006)}]{Tscherbul:2006}
\bibinfo{author}{\bibfnamefont{T.~V.} \bibnamefont{Tscherbul}}
  \bibnamefont{and} \bibinfo{author}{\bibfnamefont{R.~V.} \bibnamefont{Krems}},
  \bibinfo{journal}{Phys. Rev. Lett.} \textbf{\bibinfo{volume}{97}},
  \bibinfo{pages}{083201} (\bibinfo{year}{2006}).

\bibitem[{\citenamefont{Feshbach}(1958)}]{Feshbach:1958}
\bibinfo{author}{\bibfnamefont{H.}~\bibnamefont{Feshbach}},
  \bibinfo{journal}{Ann. Phys.} \textbf{\bibinfo{volume}{5}},
  \bibinfo{pages}{357} (\bibinfo{year}{1958}).

\bibitem[{\citenamefont{Feshbach}(1962)}]{Feshbach:1962}
\bibinfo{author}{\bibfnamefont{H.}~\bibnamefont{Feshbach}},
  \bibinfo{journal}{Ann. Phys.} \textbf{\bibinfo{volume}{19}},
  \bibinfo{pages}{287} (\bibinfo{year}{1962}).

\bibitem[{\citenamefont{Ashton et~al.}(1983)\citenamefont{Ashton, Child, and
  Hutson}}]{Ashton:1983}
\bibinfo{author}{\bibfnamefont{C.~J.} \bibnamefont{Ashton}},
  \bibinfo{author}{\bibfnamefont{M.~S.} \bibnamefont{Child}}, \bibnamefont{and}
  \bibinfo{author}{\bibfnamefont{J.~M.} \bibnamefont{Hutson}},
  \bibinfo{journal}{J. Chem. Phys.} \textbf{\bibinfo{volume}{78}},
  \bibinfo{pages}{4025} (\bibinfo{year}{1983}).

\bibitem[{\citenamefont{Hutson and {Le Roy}}(1983)}]{Hutson:HDAr:1983}
\bibinfo{author}{\bibfnamefont{J.~M.} \bibnamefont{Hutson}} \bibnamefont{and}
  \bibinfo{author}{\bibfnamefont{R.~J.} \bibnamefont{{Le Roy}}},
  \bibinfo{journal}{J. Chem. Phys.} \textbf{\bibinfo{volume}{78}},
  \bibinfo{pages}{4040} (\bibinfo{year}{1983}).

\bibitem[{\citenamefont{Bohn et~al.}(2002)\citenamefont{Bohn, Avdeenkov, and
  Deskevich}}]{Bohn:fesh:2002}
\bibinfo{author}{\bibfnamefont{J.~L.} \bibnamefont{Bohn}},
  \bibinfo{author}{\bibfnamefont{A.~V.} \bibnamefont{Avdeenkov}},
  \bibnamefont{and} \bibinfo{author}{\bibfnamefont{M.~P.}
  \bibnamefont{Deskevich}}, \bibinfo{journal}{Phys. Rev. Lett.}
  \textbf{\bibinfo{volume}{89}}, \bibinfo{pages}{203202}
  (\bibinfo{year}{2002}).

\bibitem[{\citenamefont{van~de Meerakker et~al.}(2003)\citenamefont{van~de
  Meerakker, Sartakov, Mosk, Jongma, and Meijer}}]{vandeMeerakker:2003}
\bibinfo{author}{\bibfnamefont{S.~Y.~T.} \bibnamefont{van~de Meerakker}},
  \bibinfo{author}{\bibfnamefont{B.~G.} \bibnamefont{Sartakov}},
  \bibinfo{author}{\bibfnamefont{A.~P.} \bibnamefont{Mosk}},
  \bibinfo{author}{\bibfnamefont{R.~T.} \bibnamefont{Jongma}},
  \bibnamefont{and} \bibinfo{author}{\bibfnamefont{G.}~\bibnamefont{Meijer}},
  \bibinfo{journal}{Phys. Rev. A} \textbf{\bibinfo{volume}{68}},
  \bibinfo{pages}{032508} (\bibinfo{year}{2003}).

\bibitem[{\citenamefont{Kerenskaya et~al.}(2004)\citenamefont{Kerenskaya,
  Schnupf, Heaven, and van~der Avoird}}]{Kerenskaya:2004}
\bibinfo{author}{\bibfnamefont{G.}~\bibnamefont{Kerenskaya}},
  \bibinfo{author}{\bibfnamefont{U.}~\bibnamefont{Schnupf}},
  \bibinfo{author}{\bibfnamefont{M.~C.} \bibnamefont{Heaven}},
  \bibnamefont{and} \bibinfo{author}{\bibfnamefont{A.}~\bibnamefont{van~der
  Avoird}}, \bibinfo{journal}{J. Chem. Phys.} \textbf{\bibinfo{volume}{121}},
  \bibinfo{pages}{7549} (\bibinfo{year}{2004}).

\bibitem[{\citenamefont{Sold\'{a}n and Hutson}(2004)}]{Soldan:2004}
\bibinfo{author}{\bibfnamefont{P.}~\bibnamefont{Sold\'{a}n}} \bibnamefont{and}
  \bibinfo{author}{\bibfnamefont{J.~M.} \bibnamefont{Hutson}},
  \bibinfo{journal}{Phys. Rev. Lett.} \textbf{\bibinfo{volume}{92}},
  \bibinfo{pages}{163202} (\bibinfo{year}{2004}).

\bibitem[{\citenamefont{Dhont et~al.}(2005)\citenamefont{Dhont, van Lenthe,
  Groenenboom, and van~der Avoird}}]{Dhont:2005}
\bibinfo{author}{\bibfnamefont{G.~S.~F.} \bibnamefont{Dhont}},
  \bibinfo{author}{\bibfnamefont{J.~H.} \bibnamefont{van Lenthe}},
  \bibinfo{author}{\bibfnamefont{G.~C.} \bibnamefont{Groenenboom}},
  \bibnamefont{and} \bibinfo{author}{\bibfnamefont{A.}~\bibnamefont{van~der
  Avoird}}, \bibinfo{journal}{J. Chem. Phys.} \textbf{\bibinfo{volume}{123}},
  \bibinfo{pages}{184302} (\bibinfo{year}{2005}).

\bibitem[{\citenamefont{Hutson}(2006)}]{Hutson:res:2006}
\bibinfo{author}{\bibfnamefont{J.~M.} \bibnamefont{Hutson}},
  \bibinfo{journal}{arXiv:physics/0610210}  (\bibinfo{year}{2006}).

\bibitem[{\citenamefont{Brazier et~al.}(1986)\citenamefont{Brazier, Ram, and
  Bernath}}]{Brazier:1986}
\bibinfo{author}{\bibfnamefont{C.~R.} \bibnamefont{Brazier}},
  \bibinfo{author}{\bibfnamefont{R.~S.} \bibnamefont{Ram}}, \bibnamefont{and}
  \bibinfo{author}{\bibfnamefont{P.~F.} \bibnamefont{Bernath}},
  \bibinfo{journal}{J. Mol. Spectrosc.} \textbf{\bibinfo{volume}{120}},
  \bibinfo{pages}{381} (\bibinfo{year}{1986}).

\bibitem[{\citenamefont{Mizushima}(1975)}]{Mizushima}
\bibinfo{author}{\bibfnamefont{M.}~\bibnamefont{Mizushima}},
  \emph{\bibinfo{title}{Theory of Rotating Diatomic Molecules}}
  (\bibinfo{publisher}{Wiley}, \bibinfo{address}{New York},
  \bibinfo{year}{1975}).

\bibitem[{\citenamefont{Brown and Carrington}(2003)}]{Brown:p646}
\bibinfo{author}{\bibfnamefont{J.~M.} \bibnamefont{Brown}} \bibnamefont{and}
  \bibinfo{author}{\bibfnamefont{A.}~\bibnamefont{Carrington}},
  \emph{\bibinfo{title}{Rotational Spectroscopy of Diatomic Molecules}}
  (\bibinfo{publisher}{Cambridge University Press},
  \bibinfo{address}{Cambridge}, \bibinfo{year}{2003}).

\bibitem[{\citenamefont{Hutson}(1994)}]{Hutson:CPC:1994}
\bibinfo{author}{\bibfnamefont{J.~M.} \bibnamefont{Hutson}},
  \bibinfo{journal}{Comput. Phys. Commun.} \textbf{\bibinfo{volume}{84}},
  \bibinfo{pages}{1} (\bibinfo{year}{1994}).

\bibitem[{\citenamefont{Hutson}(1993)}]{Hutson:bound:1993}
\bibinfo{author}{\bibfnamefont{J.~M.} \bibnamefont{Hutson}},
  \emph{\bibinfo{title}{Bound computer program, version 5}},
  \bibinfo{howpublished}{distributed by Collaborative Computational Project
  No.\ 6 of the UK Engineering and Physical Sciences Research Council}
  (\bibinfo{year}{1993}).

\bibitem[{\citenamefont{Johnson}(1973)}]{Johnson:1973}
\bibinfo{author}{\bibfnamefont{B.~R.} \bibnamefont{Johnson}},
  \bibinfo{journal}{J. Comput. Phys.} \textbf{\bibinfo{volume}{13}},
  \bibinfo{pages}{445} (\bibinfo{year}{1973}).

\bibitem[{\citenamefont{Johnson}(1978)}]{Johnson:1978}
\bibinfo{author}{\bibfnamefont{B.~R.} \bibnamefont{Johnson}},
  \bibinfo{journal}{J. Chem. Phys.} \textbf{\bibinfo{volume}{69}},
  \bibinfo{pages}{4678} (\bibinfo{year}{1978}).

\bibitem[{\citenamefont{Dubernet et~al.}(1991)\citenamefont{Dubernet, Flower,
  and Hutson}}]{Dubernet:1991}
\bibinfo{author}{\bibfnamefont{M.~L.} \bibnamefont{Dubernet}},
  \bibinfo{author}{\bibfnamefont{D.}~\bibnamefont{Flower}}, \bibnamefont{and}
  \bibinfo{author}{\bibfnamefont{J.~M.} \bibnamefont{Hutson}},
  \bibinfo{journal}{J. Chem. Phys.} \textbf{\bibinfo{volume}{94}},
  \bibinfo{pages}{7602} (\bibinfo{year}{1991}).

\bibitem[{\citenamefont{Hutson and Green}(1994)}]{molscat:v14}
\bibinfo{author}{\bibfnamefont{J.~M.} \bibnamefont{Hutson}} \bibnamefont{and}
  \bibinfo{author}{\bibfnamefont{S.}~\bibnamefont{Green}},
  \emph{\bibinfo{title}{Molscat computer program, version 14}},
  \bibinfo{howpublished}{distributed by Collaborative Computational Project
  No.\ 6 of the UK Engineering and Physical Sciences Research Council}
  (\bibinfo{year}{1994}).

\bibitem[{\citenamefont{Balakrishnan et~al.}(1997)\citenamefont{Balakrishnan,
  Kharchenko, Forrey, and Dalgarno}}]{Balakrishnan:scat-len:1997}
\bibinfo{author}{\bibfnamefont{N.}~\bibnamefont{Balakrishnan}},
  \bibinfo{author}{\bibfnamefont{V.}~\bibnamefont{Kharchenko}},
  \bibinfo{author}{\bibfnamefont{R.~C.} \bibnamefont{Forrey}},
  \bibnamefont{and} \bibinfo{author}{\bibfnamefont{A.}~\bibnamefont{Dalgarno}},
  \bibinfo{journal}{Chem. Phys. Lett.} \textbf{\bibinfo{volume}{280}},
  \bibinfo{pages}{5} (\bibinfo{year}{1997}).

\bibitem[{\citenamefont{Bohn and Julienne}(1997)}]{Bohn:1997}
\bibinfo{author}{\bibfnamefont{J.~L.} \bibnamefont{Bohn}} \bibnamefont{and}
  \bibinfo{author}{\bibfnamefont{P.~S.} \bibnamefont{Julienne}},
  \bibinfo{journal}{Phys. Rev. A} \textbf{\bibinfo{volume}{56}},
  \bibinfo{pages}{1486} (\bibinfo{year}{1997}).

\bibitem[{\citenamefont{Mott and Massey}(1965)}]{Mott:p380:1965}
\bibinfo{author}{\bibfnamefont{N.~F.} \bibnamefont{Mott}} \bibnamefont{and}
  \bibinfo{author}{\bibfnamefont{H.~S.~W.} \bibnamefont{Massey}},
  \emph{\bibinfo{title}{The Theory of Atomic Collisions}}
  (\bibinfo{publisher}{Clarendon Press, Oxford}, \bibinfo{year}{1965}),
  \bibinfo{edition}{3rd} ed.

\bibitem[{\citenamefont{Cvita\v{s} et~al.}(2006)\citenamefont{Cvita\v{s},
  Sold\'{a}n, Hutson, Honvault, and Launay}}]{Cvitas:li3:2006}
\bibinfo{author}{\bibfnamefont{M.~T.} \bibnamefont{Cvita\v{s}}},
  \bibinfo{author}{\bibfnamefont{P.}~\bibnamefont{Sold\'{a}n}},
  \bibinfo{author}{\bibfnamefont{J.~M.} \bibnamefont{Hutson}},
  \bibinfo{author}{\bibfnamefont{P.}~\bibnamefont{Honvault}}, \bibnamefont{and}
  \bibinfo{author}{\bibfnamefont{J.~M.} \bibnamefont{Launay}},
  \bibinfo{journal}{in preparation}  (\bibinfo{year}{2006}).

\bibitem[{\citenamefont{Timmermans et~al.}(1999)\citenamefont{Timmermans,
  Tommasini, Hussein, and Kerman}}]{Timmermans:1999}
\bibinfo{author}{\bibfnamefont{E.}~\bibnamefont{Timmermans}},
  \bibinfo{author}{\bibfnamefont{P.}~\bibnamefont{Tommasini}},
  \bibinfo{author}{\bibfnamefont{M.}~\bibnamefont{Hussein}}, \bibnamefont{and}
  \bibinfo{author}{\bibfnamefont{A.}~\bibnamefont{Kerman}},
  \bibinfo{journal}{Phys. Rep.} \textbf{\bibinfo{volume}{315}},
  \bibinfo{pages}{199} (\bibinfo{year}{1999}).

\bibitem[{\citenamefont{Moerdijk et~al.}(1995)\citenamefont{Moerdijk, Verhaar,
  and Axelsson}}]{Moerdijk:1995}
\bibinfo{author}{\bibfnamefont{A.~J.} \bibnamefont{Moerdijk}},
  \bibinfo{author}{\bibfnamefont{B.~J.} \bibnamefont{Verhaar}},
  \bibnamefont{and} \bibinfo{author}{\bibfnamefont{A.}~\bibnamefont{Axelsson}},
  \bibinfo{journal}{Phys. Rev. A} \textbf{\bibinfo{volume}{51}},
  \bibinfo{pages}{4852} (\bibinfo{year}{1995}).

\bibitem[{\citenamefont{Fano}(1961)}]{Fano:1961}
\bibinfo{author}{\bibfnamefont{U.}~\bibnamefont{Fano}}, \bibinfo{journal}{Phys.
  Rev.} \textbf{\bibinfo{volume}{124}}, \bibinfo{pages}{1866}
  (\bibinfo{year}{1961}).

\bibitem[{\citenamefont{Qu\'{e}m\'{e}ner
  et~al.}(2004)\citenamefont{Qu\'{e}m\'{e}ner, Honvault, and
  Launay}}]{Quemener:2004}
\bibinfo{author}{\bibfnamefont{G.}~\bibnamefont{Qu\'{e}m\'{e}ner}},
  \bibinfo{author}{\bibfnamefont{P.}~\bibnamefont{Honvault}}, \bibnamefont{and}
  \bibinfo{author}{\bibfnamefont{J.~M.} \bibnamefont{Launay}},
  \bibinfo{journal}{Eur. Phys. J. D} \textbf{\bibinfo{volume}{30}},
  \bibinfo{pages}{201} (\bibinfo{year}{2004}).

\end{thebibliography}
\end{document}